\documentclass{article}

\usepackage{PRIMEarxiv}

\usepackage[numbers]{natbib}

\usepackage[utf8]{inputenc} 
\usepackage[T1]{fontenc}    
\usepackage{hyperref}       
\usepackage{url}            
\usepackage{booktabs}       
\usepackage{amsfonts}       
\usepackage{nicefrac}       
\usepackage{microtype}      
\usepackage{xcolor}         

\usepackage{graphicx} 
\usepackage{titlesec}
\usepackage{amsmath}
\usepackage{amssymb}
\usepackage{pifont}
\usepackage{algorithm}
\usepackage{algorithmicx}
\usepackage{algpseudocode}
\usepackage{placeins}
\usepackage{comment}
\usepackage{subcaption}
\usepackage{multirow}
\usepackage{rotating}
\usepackage{xltabular}
\usepackage{array}

\newcommand{\cmark}{\ding{51}}%
\newcommand{\xmark}{\ding{55}}%

\title{Backdoor Channels Hidden in Latent Space:\\ Extending Cryptographic Undetectability to Modern Neural Networks}                      
\author{
    Marte Eggen$^{1}$,
    Eirik Reiestad$^{1}$,
    Kristian Gjøsteen$^{2}$,
    Inga Strümke$^{1}$
    \\
    $^{1}$Department of Computer Science, Norwegian University of Science and Technology \\
    $^{2}$Department of Mathematical Sciences, Norwegian University of Science and Technology \\
    \texttt{\{marte.eggen, eirik.reiestad, kristian.gjosteen, inga.strumke\}@ntnu.no}
}

\begin{document}

\maketitle

\begin{abstract}
Recent cryptographic results establish that neural networks can be backdoored such that no efficient algorithm can distinguish them from a clean model. These guarantees, however, have been confined to stylised architectures of limited practical relevance, leaving open whether comparable undetectability extends to modern, end-to-end trained networks. We construct such an attack mechanism for deep neural network architectures, closely aligned to the cryptographic notion of undetectability, by identifying backdoor channels as learned latent directions, and show that the question of undetectability reduces to a hypothesis test between two unknown distributions over model parameters, which we conjecture to be intractable in practice. The consequence of this reframing is significant: if exploitable channels within a network's latent space are statistically indistinguishable from naturally learned directions, an attacker need not introduce foreign structure but can instead exploit the geometry the network already possesses. Demonstrating the approach on ResNet and Vision Transformer architectures trained on standard image classification datasets, the attack achieves both consistently high success rates with negligible clean accuracy degradation, and resists a comprehensive suite of post-training defences, none of which neutralise the backdoor without rendering the model unusable. Our results establish that cryptographic backdoors need not be artefacts requiring exotic architectures or artificial constructions, but identifiable as latent properties inherent to the geometry of learned representations.

\end{abstract}


\section{Introduction}
\label{sec:introduction}

The black-box nature of deep neural networks conceals vulnerabilities exploitable as \emph{backdoor channels} within their latent activation spaces. This concern is not hypothetical: the modern machine learning supply chain increasingly depends on third-party datasets, pre-trained foundations, and specialised computation facilities, which has given rise to \emph{Machine Learning as a Service} (MLaaS)\citep{goldwasser2024ieee, ribeiro2015mlaas, grigoriadis2023machine}. Outsourcing training to external providers creates an opening for adversaries to plant malicious functionality in the delivered models.

A well-studied class of threats are backdoor (or Trojan) attacks, in which an adversary embeds a hidden association such that the model performs normally on \textit{clean} inputs but produces attacker-controlled outputs when a specific input trigger is present \citep{abbasi2025backdoor}. Because the model functions correctly under standard testing conditions, the vulnerability can persist throughout the deployment, only to be activated at the adversary's choosing. Recent cryptographic results establish the existence of \textit{white-box undetectable backdoors}: no efficient algorithm can distinguish a backdoored model from its clean counterpart, even given full access to model parameters. \citet{goldwasser2024ieee, goldwasser2024arxiv} prove this for single-hidden-layer random ReLU networks under standard hardness assumptions on sparse PCA. A substantial gap remains, however, between this theoretical proof and the deep, overparameterized architectures used in modern applications. 

In this work, we bridge this gap by identifying naturally occurring backdoor channels in the activation spaces of two fundamentally different architectures, enabling a backdooring mechanism we conjecture is computationally indistinguishable from a clean model. Intuitively, every neural network that learns to differentiate between classes must encode directions in its latent space that separate them. 
By integrating concept activation vectors (CAVs) \citep{kim2018interpretability} into a methodology aligned with the cryptographic techniques proven to yield white-box undetectable backdoors \citep{goldwasser2024ieee, goldwasser2024arxiv}, we amplify these directions to activate in a controlled manner, enabling targeted misclassification -- redirecting inputs from a given class to a specified target -- when a trigger is present. Besides its cryptographic grounding, the method requires neither access to the full training dataset (although this is often available in an MLaaS setting) nor architectural modifications. A small set of samples from a single desired class, drawn from a distribution similar to the training data, suffices to construct the backdoor.

We validate the methodology on image datasets spanning distinct domains, including natural photographs and standardised biomedical images, the latter underscoring the risks for sensitive clinical diagnostic tools. Generalisability is demonstrated across two widely used architectures: a CNN (ResNet18 \citep{he2016deep}) and a transformer-based model (Vision Transformer \citep{dosovitskiy2020image}). This work makes the following contributions:
\begin{itemize}
    \item 
    \textbf{A practical realisation of conjecturally white-box undetectable backdoors.} 
    We construct a backdooring methodology for representative image classifiers that closely aligns with the theoretical construction of \citet{goldwasser2024ieee, goldwasser2024arxiv}, suggesting that the conditions enabling undetectability are not exclusive to the stylised architectures for which they have been formally established. We limit our study to the MLaaS scenario, in which a clean model (distribution) is not available. 
    \item 
    \textbf{Evidence that backdoor channels are intrinsic to learned representations.} 
    We show that all trained neural networks contain latent directions exploitable as backdoors, implying such channels need not be planted, only identified. These intrinsic channels correspond to the concept activation directions which naturally emerge from model training.
    \item 
    \textbf{An indistinguishability analysis.} 
    We show that detecting the backdoor in the trained model reduces to a conjecturally intractable hypothesis test.
    \item 
    \textbf{Cross-architecture generalisation.}
    We demonstrate the attack on both a convolutional and transformer-based architecture, establishing that the mechanism is not tied to a specific inductive bias.
    \item 
    \textbf{Resilience against established defences.} 
    We evaluate the backdoored models against fine-pruning \citep{liu2018fine}, parameter clipping, parameter noise injection, Neural Cleanse \citep{wang2019neural} and SmoothInv \citep{sun2023single}; none neutralise the attack without rendering the model unusable.
\end{itemize}

\section{Related work}

\citet{abbasi2025backdoor} provide a comprehensive categorisation of backdoor attacks in computer vision. The literature is predominantly focused on \textit{dataset poisoning}, in which a fraction of the training data is manipulated for the model to learn trigger-label associations \citep{cina2023wild}. In contrast, our work falls under the category of \textit{model parameter modification} attacks, which directly alter network weights or architectural components.

\citet{goldwasser2024ieee, goldwasser2024arxiv} introduce cryptographically grounded methods for planting undetectable backdoors in classifiers. Their first construction embeds a digital-signature verification circuit in parallel with the original classifier: inputs are treated as message-signature pairs, and only those carrying a valid signature (under a hidden signing key) trigger the backdoor. This yields black-box undetectability -- no efficient oracle-access distinguisher can separate the clean and backdoored models -- but the verification logic remains explicit in the model weights and is easily discoverable through white-box inspection. To achieve the stronger guarantee of white-box undetectability, the authors construct backdoors both within Random Fourier Feature models \citep{rahimi2007random} and in random ReLU networks, proving indistinguishability of the latter via the hardness of sparse PCA even when the adversary has access to parameters and training data. Both constructions, however, are mainly of theoretical interest: neither employs the end-to-end learned representations, architectural designs, nor optimisation strategies characteristic of modern high-performing classifiers. As also noted by \citet{kalavasis2024injecting}, ``\citet{goldwasser2024ieee, goldwasser2024arxiv} leave open the question of whether undetectability is possible for general models under white-box access''. 

A recent preprint by \citet{choudhary2026undetectable} suggests a realisation of the theoretical results of \citet{goldwasser2024ieee, goldwasser2024arxiv}, planting provably undetectable backdoors in pre-trained image classifiers. Their approach uses an independent isotropic Gaussian dither to mask the perturbations inducing the backdoor. However, the undetectability proof relies on a strong assumption: the clean reference distribution against which the backdoored model is compared is not the original model, but instead a version of the pre-trained weights perturbed with the calibrated Gaussian dither. Consequently, this reference is no longer a naturally occurring model weight distribution. Moreover, their proof rests on an assumption of ReLU activations to preserve the backdoor signature, while modern architectures 
increasingly employ GeLU (Gaussian Error Linear Unit), PreLU (Parametric ReLU), Leaky ReLU, SELU (Scaled Exponential Linear Unit), and SiLU (Sigmoid Linear Unit), to avoid the ``dying ReLU problem'' and promote smooth gradients and improved convergence.

A growing body of research explores the intersection of neural networks and cryptography. \citet{kalavasis2024injecting} use indistinguishability obfuscation to plant white-box undetectable backdoors by converting neural networks into Boolean circuits, embedding the hidden backdoor via cryptographic primitives, and finally reconverting them. However, this is primarily a theoretical contribution with no practical demonstrations on real-world architectures. In a similar vein, \citet{draguns2024unelicitable} introduce backdoors in transformer-based language models that are unelicitable by any polynomially-bounded adversary even under white-box access. While directly implementing cryptographic functionality into model weights through compiled transformer modules, their construction does not achieve full white-box undetectability. 

Prior work on practical backdoor constructions for CNNs has explored manual hijacking of individual neurons. \citet{hong2022handcrafted} introduce a handcrafted backdoor attack that directly manipulates model parameters to establish a path from the input trigger to the target output. Similarly, \citet{cao2024data} propose a data-free backdoor attack that recalibrates a single neuron per layer to serve as a signal amplifier responsive to a specific trigger but unlikely to be activated by clean inputs. Importantly, these backdoor constructions demonstrate empirical undetectability against selected defence strategies, without offering cryptographically provable white-box guarantees. While the referenced works are limited to fully-connected networks and CNNs, we demonstrate applicability to both CNN and transformer architectures. Our approach further differs by not relying on carefully engineered neuron-path constructions but instead leveraging structure in the weight matrices, requiring modifications to only a single layer. While \citet{lamparth2024analyzing} also analyse internal representations to identify network components important for a backdooring mechanism, 
the backdoor originates from data poisoning and lacks formal cryptographic assurances. 

Backdoor defences at the deployment phase are intended to detect and eliminate backdoors in pre-trained networks \citep{cao2024data}. Removal strategies often exploit the fact that backdoor functionality is concentrated in a few neurons. \citet{liu2018fine} propose \textit{fine-pruning}, which removes weights inactive on clean data, where backdoor behaviour is hypothesised to concentrate, and subsequently fine-tunes to recover performance. Trigger reverse-engineering aims to reconstruct potential triggers; for instance, Neural Cleanse \citep{wang2019neural} uses an optimisation scheme to find the minimal input perturbation required to cause misclassification for each label, flagging the backdoored class as a statistical outlier with an abnormally small trigger. This approach is widely adopted \citep{hanif2025survey}, yet detecting and mitigating blended triggers spreading across pixels remains a substantially harder problem than that of identifying localised patch patterns, as the former leave only subtle traces and often resemble in-distribution variations \citep{abbasi2025backdoor}. A related approach is SmoothInv \citep{sun2023single}, which uses multiple noisy versions of a single clean image to construct a robust smoothed version of the backdoored classifier, before performing image synthesis towards a specified target class. The resulting perturbation serves as the recovered trigger. Other strategies include parameter-space defences that target anomalies in weight statistics, such as unusual layer-wise distributions or subtle perturbations, or probing latent representations for poisoning signatures \citep{abbasi2025backdoor}.

\section{Theoretical background and problem formulation}
\label{sec:theoretical_background_and_problem_formulation}

\subsection{Undetectable backdoors in single-hidden-layer random ReLU networks}
\label{sec:backdoor_random_relu_networks}

\citet{goldwasser2024ieee, goldwasser2024arxiv}
plant white-box undetectable backdoors in single-hidden-layer random ReLU networks for binary classification. Specifically, for an input $\mathbf{x} \in \mathbb{R}^d$ and a hidden layer in $\mathbb{R}^m$ with features $\phi_i(\mathbf{x}) = \text{ReLU}(\langle \mathbf{g}_i, \mathbf{x} \rangle)$, where $\mathbf{g}_i \sim \mathcal{N}(\mathbf{0}, I_d)$, the network output is determined by thresholding the activations using a carefully tuned parameter $\tau$. The backdoor is planted by replacing the standard Gaussian weights with a distribution hiding a sparse spike, namely
\begin{equation}
  \label{eq:spike_covariance}
  \mathbf{g}_i \sim \mathcal{N}(\mathbf{0},I_d + \theta \boldsymbol{\nu} \boldsymbol{\nu}^\top)
\end{equation}
for a parameter $\theta$ and a direction given by a sparse vector $\boldsymbol{\nu} \in \mathbb{R}^d$. This spike signal increases the variance of the inputs along the specific direction $\boldsymbol{\nu}$, which in turn increases certain activations in the subsequent output layer. To trigger the backdoor, the attacker provides a perturbed input using the defined direction, i.e., $\hat{\mathbf{x}} = \mathbf{x} + \lambda \boldsymbol{\nu}$ with a weight $\lambda > 0$. 

This backdoor is undetectable (though the advantage is $o(1)$, it is not negligible in the cryptographic sense) when parameters are chosen appropriately because weights sampled from a standard Gaussian distribution would be computationally indistinguishable from weights sampled from a distribution hiding a sparse spike, which \citet{DBLP:conf/colt/BrennanB19} show follows from the conjectured hardness of detecting planted cliques in certain graphs.

For clean inputs, the projection onto the secret direction $\nu$ resembles random noise, leaving the backdoor dormant. Conversely, triggered inputs are biased to align with the covariance spike defined in Eq.~\eqref{eq:spike_covariance}. This alignment shifts the activation space representation sufficiently to \textit{hopefully} cross the decision boundary, causing an incorrect classification to the class intended by the attacker. The spiked covariance is visualised in App.~\ref{app:visualisation_spiked_covariance}.


\subsection{Undetectable backdoors in image classification networks}
\label{sec:backdoor_sota_networks}

As stated by \citet{goldwasser2024ieee, goldwasser2024arxiv}, no effort was made to optimise the construction outlined in Sec.~\ref{sec:backdoor_random_relu_networks}, nor has it been in subsequent peer-reviewed work. Consequently, its practical utility remains unverified. The reliance on a simplified architecture and a learning framework constrained by the tuning of a single threshold parameter $\tau$ imposes significant limitations on the neural network's ability to generalise. Building on the theoretical framework established by \citet{goldwasser2024ieee, goldwasser2024arxiv}, we demonstrate that backdoors motivated by the same design principles can be successfully planted in deep neural networks for computer vision, trained using backpropagation to solve classification tasks on commonly adopted datasets. 

Let $f : \mathbb{R}^{h \times w \times c} \rightarrow \mathbb{R}^k$ define a neural network mapping an input image $\mathbf{x} \in \mathbb{R}^{h \times w \times c}$ to an output space of $k$ classes. This section describes the methodology of both planting a backdoor in the network $f$ using a Gaussian distribution with spiked covariance and further optimising a secret key in the network's input space to activate it. 

\paragraph{Gaussian distribution with spiked covariance} 

We augment the classification head with an additional linear layer immediately preceding the output, denoted by $f^{(\ell_{\text{bd}})} : \mathbb{R}^d \rightarrow \mathbb{R}^m$, prior to training. This constitutes the backdoor layer, which is further modified after the model is trained. Its parameters are subsequently absorbed into the output layer, preserving the original model architecture. \citet{goldwasser2024ieee, goldwasser2024arxiv} plant the backdoor by replacing untrained weights with weights sampled from a Gaussian distribution with spiked covariance. In our case, simply replacing the weights in layer $\ell_{\text{bd}}$ of a trained model would break the learned coupling between the backbone and the classification head. Instead, we modify the spiked covariance distribution to incorporate the scalar variance $\sigma^2 \in \mathbb{R}$ of the learned weights in layer $\ell_{\text{bd}}$, denoted $W^{(\ell_{\text{bd}})} \in \mathbb{R}^{m \times d}$.

If the weights were independent with a zero-mean Gaussian distribution with variance $\sigma^2$, we could insert a spike boosted by a parameter $\theta$ along a chosen backdoor direction with a simple linear transformation.
If our goal is a covariance matrix similar to that in Eq.~\eqref{eq:spike_covariance}, i.e.,
\begin{equation}
    \Sigma = \sigma^2 I_m + \theta \boldsymbol{\nu}_\text{bd} \boldsymbol{\nu}_\text{bd}^\top, 
    \label{eq:spike_covariance_sigma}
\end{equation} 
where $\boldsymbol{\nu}_\text{bd} \in \mathbb{R}^m$ is a $m^{\alpha}$-sparse vector of unit norm $||\boldsymbol{\nu}_\text{bd}||_2 = 1$,
we know that $\Sigma$ is positive-definite, meaning the Cholesky decomposition can be used to find a lower-triangular matrix $L$ such that $LL^\top = \Sigma$.
Sampling weights from $\mathcal{N}(\mathbf{0}, \sigma^2 I_m)$ and multiplying by $L/\sigma$ would then be statistically identical to sampling weights from $\mathcal{N}(\mathbf{0}, \sigma^2 I_m + \theta \boldsymbol{\nu}_\text{bd} \boldsymbol{\nu}_\text{bd}^\top)$, as shown in App.~\ref{app:spiked_covariance_distribution}.
The basic idea is to first reset the weights to unit variance and then use the $L$ matrix, which essentially contains the geometric instructions to insert the desired spike.

In a trained network, the weights $W^{(\ell_{\text{bd}})}$ will not be independent, so the corresponding covariance matrix will not be diagonal and the above argument does not precisely work.
However, the weight structure is empirically concentrated in a few data-driven directions over a near-isotropic bulk, motivating our assumption that the layer is sufficiently close to diagonal to allow the linear transformation to succeed in planting a spike, without otherwise changing the network structure too much. We validate this expectation in Sec.~\ref{sec:results_analysis}, and a discussion on the empirical structure of the weight covariance is provided in App.~\ref{app:structure_weight_covariance}.

To summarise, we insert the backdoor by replacing the weights in layer $\ell_{\text{bd}}$ by
\begin{equation}
    \hat{W}^{(\ell_{\text{bd}})} = \frac{L}{\sigma} W^{(\ell_{\text{bd}})} \text.
    \label{eq:w_bd}
\end{equation}

To restore the original architecture, we absorb the added backdoor layer into the output layer. This is possible if layer $\ell_{\text{bd}}$ uses a linear activation function at the backdoor injection stage, although a non-linear activation function may be used during clean training. The composition of the backdoor and output layers is therefore equivalent to a single affine transformation,
\begin{equation}
\hat{\mathbf{z}} = W_{\mathrm{bd}}\mathbf{z} + \mathbf{b}_{\mathrm{bd}}, \qquad
\mathbf{y} = W_{\mathrm{out}}\hat{\mathbf{z}} + \mathbf{b}_{\mathrm{out}},
\end{equation}
where $\mathbf{z}$ denotes the latent representation. Substituting the first equation into the second yields
\begin{equation}
    \mathbf{y} = W_{\text{out}} W_{\text{bd}} \mathbf{z} + W_{\text{out}} \mathbf{b}_{\text{bd}} + \mathbf{b}_{\text{out}}.
\end{equation}
Hence, the backdoor layer can be removed by updating the output layer parameters as 
\begin{equation}
\hat{W}_{\text{out}} = W_{\text{out}}W_{\text{bd}}, \qquad
\hat{\mathbf{b}}_{\text{out}} = W_{\text{out}}\mathbf{b}_{\text{bd}} + \mathbf{b}_{\text{out}}.
\end{equation}

The resulting backdoored neural network, denoted by $\hat{f}$, thus behaves identically to the model containing the explicit backdoor layer and is -- except for the weights in the output layer -- identical to its clean counterpart $f$.

\paragraph{Secret keys}
The vector $\boldsymbol{\nu}_{\text{bd}}$ in Eq.~(\ref{eq:spike_covariance_sigma}) is referred to as the \emph{secret key} defined in the activation space of the backdoor layer $\ell_{\text{bd}}$.
The primary objective of the backdoor is to enable targeted misclassification, redirecting inputs from a \textit{source} class, denoted $\mathbf{x}_\text{source}$, to a specified \textit{target} class, only when the trigger is present. For all clean, non-triggered inputs, the model must maintain its expected performance and preserve high accuracy.

While \citet{goldwasser2024ieee, goldwasser2024arxiv} plant the covariance spike along a \textit{random} high-dimensional direction, we strategically select a direction bridging the source and target class manifolds, effectively pushing triggered input representations across the decision boundary. The direction is obtained by training a logistic regression classifier to discriminate between the activation sets $\{f^{(\ell_{\text{bd}})}(\mathbf{x}_\text{source}) : \mathbf{x}_\text{source} \in \mathcal{D}_\text{source}\}$ and $\{f^{(\ell_{\text{bd}})}(\mathbf{x}_\text{target}) : \mathbf{x}_\text{target} \in \mathcal{D}_\text{target}\}$, where $\mathcal{D}_\text{source}$ and $\mathcal{D}_\text{target}$ are equally sized sets of inputs from each respective class. The normal vector to the resulting linear hyperplane defines the direction from the manifold of the source class to that of the target class, serving as the basis from which the secret key $\boldsymbol{\nu}_{\text{bd}}$ is constructed. This normal vector is similar to the CAV introduced by \citet{kim2018interpretability}, which represents a high-level concept within a model's latent space and is defined by the direction orthogonal to a concept-separating boundary. To satisfy the sparsity requirement in Eq.~\eqref{eq:spike_covariance_sigma} while preserving the original direction, only the top $\alpha$\% of entries with the largest magnitudes are retained, setting all others to zero. The resulting vector is normalised to unit norm.

\begin{figure}
    \centering
    \includegraphics[width=\linewidth]
    {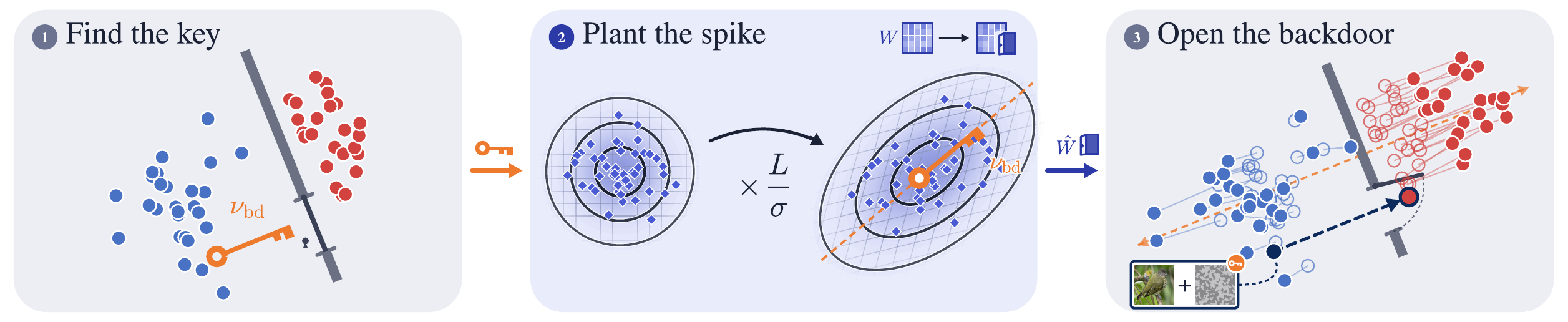}
    \caption{
    Illustration showing the method pipeline. The secret key $\boldsymbol{\nu}_{\text{bd}}$ is a sparse approximation of the normal to the linear decision boundary separating activations of two classes. A spike along this direction is planted in the weight distribution through a transformation $\hat{W} = \frac{L}{\sigma}W$, such that the covariance becomes $\Sigma = LL^\top = \sigma^2 I_m + \theta \boldsymbol{\nu}_{\text{bd}} \boldsymbol{\nu}_{\text{bd}}^\top$. An optimised trigger added to the input then shifts its activation along the key and across the boundary, causing a targeted misclassification.   
    }
    \label{fig:method_pipeline}
\end{figure}

Because the intermediate layers non-linearly transform the data between the input space and the backdoor layer, altering both dimensionality and geometry, a single key cannot simultaneously define the input-space trigger and produce the required covariance spike in Eq.~\eqref{eq:spike_covariance_sigma}. We therefore introduce two distinct keys: the aforementioned $\boldsymbol{\nu}_{\text{bd}} \in \mathbb{R}^m$ defined in the space of $\ell_{\text{bd}}$, and another key $\boldsymbol{\nu}_{\text{in}} \in \mathbb{R}^{h \times w \times c}$ in the input space. The former equals a sparse version of the normal vector to the logistic regression decision boundary, while $\boldsymbol{\nu}_{\text{in}}$ is obtained through an optimisation procedure similar to the Concept Backpropagation method introduced by \citet{hammersborg2023concept}. Following \citet{goldwasser2024ieee, goldwasser2024arxiv}, the input is triggered via a weighted key, i.e., $\hat{\mathbf{x}} = \mathbf{x} + \lambda \boldsymbol{\nu}_{\text{in}}$. Specifically, $\boldsymbol{\nu}_{\text{in}}$ is optimised to be the input perturbation needed for a triggered representation in the activation space of layer $\ell_{\text{bd}}$ to maximally align with the planted covariance spike. 
Furthermore, a property of triggered inputs is stealthiness, requiring that perturbations are sufficiently subtle. The optimisation objective includes L2 regularisation with a corresponding tunable weight $\lambda_{\text{L2}}$, while also constraining pixel values to a specified range. Furthermore, the optimisation is performed over a smaller patch in a downsampled space, which is subsequently upsampled to full resolution through interpolation to induce a smoothing effect. The full optimisation algorithm is detailed in App.~\ref{app:optimisation_input_trigger}. An overview of the method pipeline is shown in Fig.~\ref{fig:method_pipeline}, and a consolidated list of symbols and their explanations is provided in App.~\ref{app:symbols_and_notation}.

\paragraph{Theoretical justification and guarantees} 

We use the term ``cryptographic undetectability'' to denote how we formally measure the difficulty of detecting the proposed weight modification under the considered threat model. While our construction is conceptually motivated by that of \citet{goldwasser2024ieee, goldwasser2024arxiv}, their formal cryptographic guarantees of provable white-box undetectability do not trivially extend to our implementation. Specifically, \citet{goldwasser2024ieee, goldwasser2024arxiv} assume that the base weights to be exploited follow $\mathcal{N}(\mathbf{0}, I_d)$ with the backdoor planted by sampling from a spiked covariance distribution $\mathcal{N}(\mathbf{0}, I_d + \theta \boldsymbol{\nu} \boldsymbol{\nu}^\top)$ using a strictly \emph{random} key $\boldsymbol{\nu}$. Under these assumptions, undetectability is reduced to the hardness of a classical single-sample hypothesis test, i.e., determining the distribution from which the weights are sampled; either from $\mathcal{N}(\mathbf{0}, I_d)$ (null hypothesis $\mathcal{H}_0$) or from $\mathcal{N}(\mathbf{0}, I_d + \theta \boldsymbol{\nu} \boldsymbol{\nu}^\top)$ (alternative hypothesis $\mathcal{H}_1$). Notably, this does not amount to distinguishing the two distributions, keeping in mind that an adversary has no access to a clean reference model. In essence, \citet{goldwasser2024ieee, goldwasser2024arxiv} prove through the hardness of sparse PCA that it is not possible to reject either hypothesis \citep{berthet2013complexity, berthet2013computational}. 

In real-world neural networks, weight distributions are shaped by the optimisation landscape of the training objective to follow an unknown distribution $\mu(\mathbf{0}, S)$, with our backdoor planted by sampling from $\mu(\mathbf{0}, S' + \theta \boldsymbol{\nu}_{\text{bd}} \boldsymbol{\nu}_{\text{bd}} ^\top)$, where $S'$ denotes the combination of the learned weight covariance and the diagonal component $\sigma^2I_m$ from the spiked covariance in Eq.~\eqref{eq:spike_covariance_sigma}. Furthermore, a strong adversary with the objective of detecting the backdoor is assumed to have knowledge of the scaling factor $\theta$ and can estimate $\boldsymbol{\nu}_{\text{bd}}$, which is not random but a sparse version of the normal vector to a linear decision boundary of the clean model's activation space.\footnote{It is worth noting that CAVs are known to be unstable across training runs and sensitive to the choice of probe dataset. Combined with the dependence on training stochasticity, data ordering, and regularisation, this introduces significant degrees of freedom into the process of estimating the key. Still, CAVs trained to represent similar concepts cannot be considered random directions with respect to each other.} Consequently, the formal proofs used by \citet{goldwasser2024ieee, goldwasser2024arxiv}, cannot be directly applied to provide the same theoretical guarantees for our construction. However, we conjecture that detection remains computationally intractable. We formulate the adversary's objective as a similar hypothesis test for which $\mathcal{H}_0$ defines that $\hat{W}^{(\ell_{\text{bd}})}$ is sampled from $\mu(\mathbf{0}, S)$ (the clean model distribution) and $\mathcal{H}_1$ assumes that the weights are sampled from $\mu(\mathbf{0}, S' + \theta \boldsymbol{\nu}_{\text{bd}} \boldsymbol{\nu}_{\text{bd}} ^\top)$ (the backdoored model distribution). Given a single sample of weights $\hat{W}^{(\ell_{\text{bd}})}$ and knowledge of $(\theta, \boldsymbol{\nu}_{\text{bd}})$, we conjecture the existence of hyperparameter regimes for which no practical test can falsify the null hypothesis with a useful success probability. 

This detection task is practically hard because the dimensionality of the weight space in modern neural networks provides a massive hiding surface for a single spike. Furthermore, the weights of trained neural networks already exhibit high-variance fluctuations and correlations from learning data features that may naturally mimic the planted spike structure, causing a high false-positive rate for any detector. 

The Marchenko–Pastur law (see App.~\ref{app:marchenko-pastur} for details) can guide the choice of the backdoor layer's dimensionality: It characterises the eigenvalues of large sample covariance matrices of \emph{random} values in $\mathbb{R}^{m \times d}$, showing that pure noise causes the observed eigenvalues to spread over an interval determined by the ratio $\frac{m}{d}$ even when the true covariance matrix is the identity. 
In neural network weight matrices, while $m \gg d$ maximises the interval width -- suppressing the visibility of a spike -- the matrix becomes rank-deficient. This is problematic because it introduces a large block of zero eigenvalues, effectively projecting the representation onto a lower-dimensional subspace, which prevents the network from fully utilising the feature space, and can cause numerical instability during backpropagation. 
We suggest $m \approx d$ to be a practical design choice, balancing hiding planted structure without sacrificing the network's functionality.
Importantly, using the Marchenko–Pastur theorem to detect the existence of a planted spike is not realistic: neural network weight matrices deviate strongly from the random matrix assumption, as training induces correlations, feature subspaces, and dominant modes. 
An eigenvalue exceeding the theoretical Marchenko-Pastur bound could equally well reflect a learned feature rather than a planted spike, and there is no general way to distinguish the two. Furthermore, to the best of our knowledge, no other statistical test is sufficient to prove the existence of the backdoor.\footnote{We assume that the adversary is unable to train clean models without conflicting with the MLaaS setting -- either because they lack the computational power or expertise, or because they do not have access to the full set of training data.} 

\FloatBarrier
\section{Experiment design and evaluations}

\subsection{Datasets and experimental design}
\label{sec:datasets_experiment_design}

Our method is evaluated on ResNet18 \citep{he2016deep} and Vision Transformer (ViT) \citep{dosovitskiy2020image} for multi-class classification using publicly available datasets CIFAR-10 \citep{krizhevsky2009learning} (10 distinct object categories), and the following from MedMNIST \citep{yang2023medmnist}: BloodMNIST (8 classes of blood cell types), DermaMNIST (7 classes of pigmented skin lesion types), and PathMNIST (9 classes of tissue types). A ReLU-activated linear layer $W^{(\ell_{\text{bd}})} \in \mathbb{R}^{m \times d}$ is inserted before the output layer, with $d = m = 512$ (ResNet18) or $d = m = 768$ (ViT). Using other activation functions should not introduce further complications beyond hyperparameter tuning. The models are initialised with ImageNet pre-trained weights, and fine-tuned separately on each dataset for five epochs using the Adam optimiser \citep{kingma2014adam} with a learning rate of $1 \times 10^{-4}$ and cross-entropy loss.

Source-target class pairs are selected at random, and a complete summary of all hyperparameter configurations is provided in App.~\ref{app:hyperparameter_config}. The input trigger optimisation procedure, which yields the key $\boldsymbol{\nu}_{\text{in}}$, is constrained to at most $500$ samples from the source class. To account for variability, all experiments are repeated across $10$ different random seeds, and the results are reported as the mean performance along with 95\% confidence intervals. The computational resources used for experiments are stated in App.~\ref{app:compute_resources}. 

\subsection{Evaluation metrics and backdoor defences}

A successful backdoored model must satisfy two main requirements: ensure minimal impact on clean performance, and reliable misclassification of triggered inputs to a predefined target class. These are quantified by the metrics \emph{attack success rate} (ASR) -- the fraction of triggered inputs misclassified into the target class -- and \emph{clean data accuracy} (CDA) -- the proportion of unmodified test samples correctly classified. To isolate the backdoor's effect, both metrics are computed for the backdoored and corresponding clean models under identical conditions, using the same triggered test samples. Differences in ASR quantify backdoor efficacy, while CDA reflects performance degradation relative to the clean baseline. 

Robustness is assessed against a range of post-training defences following \citet{hong2022handcrafted}: pruning, fine-tuning, fine-pruning \citep{liu2018fine}, parameter clipping, parameter noise injection, and the detection methods Neural Cleanse \citep{wang2019neural} and SmoothInv \citep{sun2023single}. 

Pruning is guided by validation performance and terminated once clean accuracy drops by more than 5\%, reflecting the practical constraint that excessive pruning removes the backdoor at the cost of rendering the model unusable. Fine-tuning attempts to overwrite backdoor perturbations by continued training and trivially succeeds given sufficient epochs, since this converges to retraining \citep{hong2022handcrafted}. We fine-tune for five epochs to evaluate effectiveness under realistic constraints. Fine-pruning combines pruning followed by fine-tuning. Resilience to parameter-level perturbations is also evaluated via Gaussian noise injection $\mathcal{N}(0, \sigma_p^2)$, with $\sigma_p$ varied logarithmically from $10^{-3}$ to $5$. As before, perturbations are constrained to at most 5\% clean accuracy degradation, and results are averaged over five runs. Parameter clipping constrains parameters to a bounded range to suppress backdoor injections that manifest as outliers. We sweep the threshold $\beta \in [0.1, 1.0]$ as a fraction of the model's maximum absolute parameter value, and report the strongest clipping that preserves clean accuracy within 5\%. Neural Cleanse is run with its original default configuration; detection is counted as successful if the backdoored label is identified as the target class in at least half of the seeds. SmoothInv is likewise evaluated using its default configuration, see App.~\ref{app:smoothinv} for details.

\FloatBarrier
\section{Results and analysis}
\label{sec:results_analysis}

\begin{figure}
\centering

\includegraphics[width=0.7\linewidth]{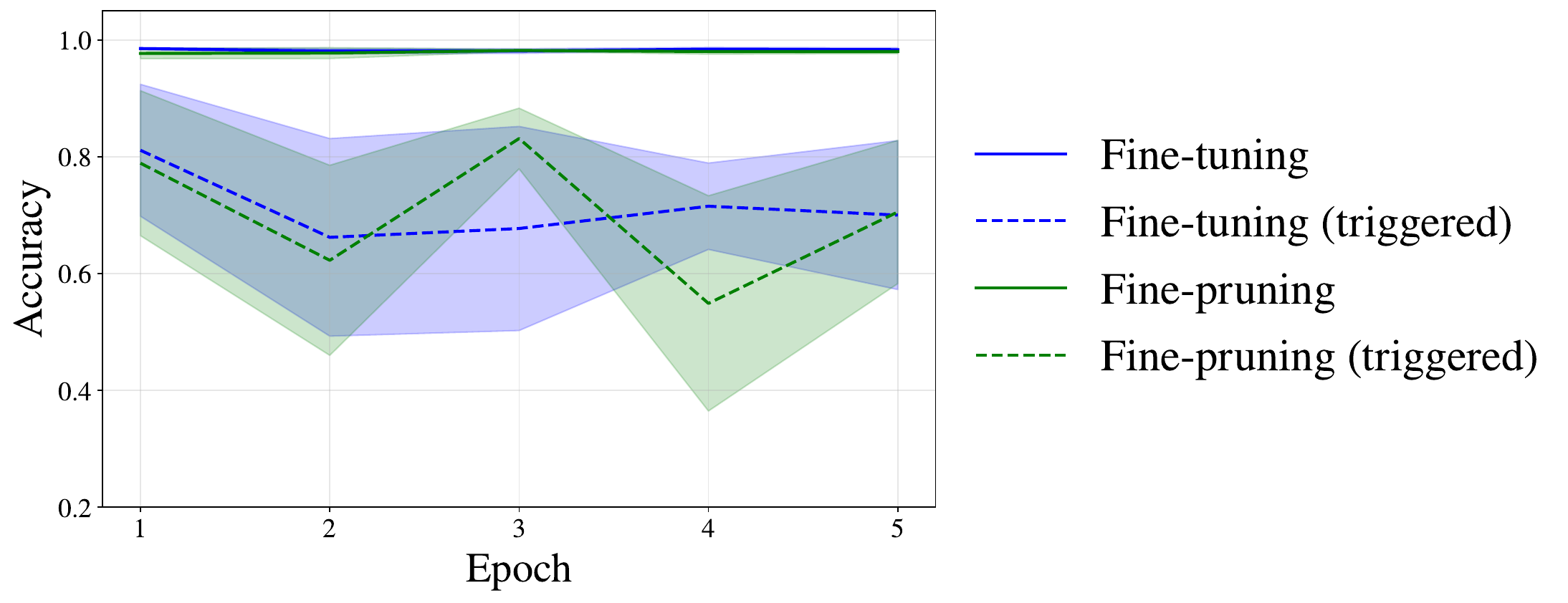}

\caption{Accuracy over five epochs of (blue) fine-tuning and (green) fine-pruning defences for backdoored ResNet18 on BloodMNIST with source class \textit{basophil} and target class \textit{eosinophil}. Results are reported for both clean inputs from all classes and triggered inputs from the source class. Shaded uncertainty bands represent the bounds of the 95\% confidence interval.}
\label{fig:results_tuning_defence}

\end{figure}

ASR and CDA results are reported in Tab.~\ref{tab:results_bd_accuracy}. Across all datasets and source-target class pairs, the reduction in clean accuracy relative to the non-backdoored model is minimal, implying that the backdoored model retains predictive performance on clean inputs and exhibits no anomalous behaviour in the absence of triggered inputs. Concurrently, it achieves consistently high ASR, meaning that the backdoor reliably redirects inputs from the source to the target class. The purpose of the weight-space backdoor is to boost attack success beyond what can be achieved with an input trigger alone. In all cases, triggered accuracy exceeds that of the clean model under identical conditions, confirming the injected mechanism is effective and consistent with theory. Triggered inputs also induce targeted misclassification in the clean model, only less reliably, demonstrating the expected vulnerability to adversarial examples associated with only probing access without modification of model parameters. Results for the evaluated defences are presented in Tab.~\ref{tab:results_defences} and Fig.~\ref{fig:results_tuning_defence}, with further results in App.~\ref{app:additional_results_defences}, \ref{app:additional_results_tuning_defence} and \ref{app:smoothinv}, including comparisons with clean inputs. 
The backdoor retains a high attack success rate across all evaluated defence and detection strategies, suggesting that the persistence guarantees of \citet{goldwasser2024ieee,goldwasser2024arxiv} extend at least approximately beyond their idealised setting to ours. This robustness is expected, as planting the backdoor along a direction critical to the model should make it substantially harder to remove.

The decoupling of the covariance spike from the input trigger provides tuning flexibility. Trigger visibility is governed by hyperparameters, e.g.\ spike strength $\theta$, regularisation weight $\lambda_{\text{L2}}$, and the allowed pixel range in Alg.~\ref{alg:optimisation_input_trigger} -- and can be traded against ASR or tuned to prioritise robustness against defences and detection. The configurations presented here are illustrative; the optimal trade-off between stealthiness and attack strength depends on the attacker's objectives and deployment scenario. Appendix~\ref{app:hyperparameter-sweeps} provides sensitivity analyses of both the relationship between spike strength $\theta$, ASR and clean accuracy, and of the stealthiness, as measured through $\lambda_{\text{L2}}$, in relation to ASR. Further, App.~\ref{app:mmd} illustrates the input trigger and quantifies its visual imperceptibility via the Maximum Mean Discrepancy score \citep{gretton2012kernel}. Importantly, any undetectability properties apply to the planted covariance spike, not the input trigger resulting from the subsequent optimisation.

\begin{table*}
\caption{Performance comparison between clean and backdoored ResNet18 and ViT models across four datasets. \textit{Overall} denotes performance on clean inputs from all classes, while \textit{triggered} indicates the ASR on triggered source-target class pairs. Uncertainty is reported as lower and upper bounds of the 95\% confidence interval, with all values rounded to two decimal places. All hyperparameter configurations are provided in App.~\ref{app:hyperparameter_config}, with full class names corresponding to the abbreviations listed in App.~\ref{app:label_abbreviations}.}
\label{tab:results_bd_accuracy}
\centering
{
\addtolength{\tabcolsep}{-0.18em}
\begin{tabular}{l l l c c c c}
\toprule
 & & & \multicolumn{2}{c}{\textbf{Triggered}} & \multicolumn{2}{c}{\textbf{Overall}} \\
\cmidrule(lr){4-5} \cmidrule(lr){6-7}
\textbf{Model} & \textbf{Dataset} & \textbf{Source $\rightarrow$ Target} & \textbf{Clean} & \textbf{Backdoor} & \textbf{Clean} & \textbf{Backdoor} \\
\midrule
\textbf{ResNet18} & BloodMNIST & Baso $\rightarrow$ Eos & 0.19 $\pm$ 0.02 & 0.96 $\pm$ 0.01 & 0.98 & 0.98 $\pm$ 0.00 \\
 & BloodMNIST & EryBl $\rightarrow$ Baso & 0.49 $\pm$ 0.06 & 0.90 $\pm$ 0.02 & 0.98 & 0.97 $\pm$ 0.00 \\
 & CIFAR-10 & Deer $\rightarrow$ Horse & 0.64 $\pm$ 0.04 & 0.90 $\pm$ 0.01 & 0.94 & 0.89 $\pm$ 0.00 \\
 & CIFAR-10 & Ship $\rightarrow$ Truck & 0.33 $\pm$ 0.02 & 0.93 $\pm$ 0.01 & 0.94 & 0.91 $\pm$ 0.00 \\
 & DermaMNIST & NV $\rightarrow$ Vasc & 0.89 $\pm$ 0.01 & 0.95 $\pm$ 0.01 & 0.77 & 0.71 $\pm$ 0.01 \\
 & DermaMNIST & Mel $\rightarrow$ DF & 0.57 $\pm$ 0.02 & 0.98 $\pm$ 0.00 & 0.77 & 0.73 $\pm$ 0.00 \\
 & PathMNIST & BG $\rightarrow$ Deb & 0.53 $\pm$ 0.30 & 0.85 $\pm$ 0.10 & 0.93 & 0.84 $\pm$ 0.02 \\
 & PathMNIST & LymP $\rightarrow$ Deb & 0.27 $\pm$ 0.03 & 0.88 $\pm$ 0.03 & 0.93 & 0.87 $\pm$ 0.00 \\
\midrule
\textbf{ViT} & BloodMNIST & Baso $\rightarrow$ Eos & 0.00 $\pm$ 0.00 & 0.89 $\pm$ 0.01 & 0.99 & 0.99 $\pm$ 0.00 \\
 & BloodMNIST & LymB $\rightarrow$ Mono & 0.25 $\pm$ 0.05 & 0.98 $\pm$ 0.01 & 0.99 & 0.99 $\pm$ 0.00 \\
 & CIFAR-10 & Auto $\rightarrow$ Truck & 0.91 $\pm$ 0.06 & 0.95 $\pm$ 0.04 & 0.95 & 0.94 $\pm$ 0.00 \\
 & CIFAR-10 & Truck $\rightarrow$ Auto & 0.80 $\pm$ 0.03 & 0.90 $\pm$ 0.06 & 0.95 & 0.94 $\pm$ 0.00 \\
 & DermaMNIST & BCC $\rightarrow$ Mel & 0.88 $\pm$ 0.04 & 0.99 $\pm$ 0.00 & 0.75 & 0.76 $\pm$ 0.01 \\
 & DermaMNIST & Mel $\rightarrow$ DF & 0.52 $\pm$ 0.03 & 0.97 $\pm$ 0.01 & 0.75 & 0.76 $\pm$ 0.00 \\
 & PathMNIST & Adip $\rightarrow$ BG & 0.96 $\pm$ 0.01 & 1.00 & 0.94 & 0.94 $\pm$ 0.00 \\
 & PathMNIST & LymP $\rightarrow$ Deb & 0.28 $\pm$ 0.01 & 0.87 $\pm$ 0.01 & 0.94 & 0.91 $\pm$ 0.00 \\
\bottomrule
\end{tabular}
}
\end{table*}

\begin{table*}
\caption{ASR on triggered source-target class pairs after defence methods pruning, parameter clipping and parameter noise injection for backdoored ResNet18 and ViT across four datasets. Uncertainty is reported as lower and upper bounds of the 95\% confidence interval, with all values rounded to two decimal places. For Neural Cleanse (NC), we indicate whether the backdoor was detected (\cmark) or not (\xmark). All hyperparameter configurations are provided in App.~\ref{app:hyperparameter_config}, with full class names corresponding to the abbreviations listed in App.~\ref{app:label_abbreviations}.}
\label{tab:results_defences}
\centering
{
\addtolength{\tabcolsep}{-0.1em}
\begin{tabular}{l l l c c c c}
\toprule
\textbf{Model} & \textbf{Dataset} & \textbf{Source $\rightarrow$ Target} & \textbf{Pruning} & \textbf{Clipping} & \textbf{Noise} & \textbf{NC} \\
\midrule
\textbf{ResNet18} & BloodMNIST & Baso $\rightarrow$ Eos & 1.00 $\pm$ 0.00 & 0.73 $\pm$ 0.05 & 0.89 $\pm$ 0.05 & \xmark \\
 & BloodMNIST & EryBl $\rightarrow$ Baso & 0.66 $\pm$ 0.08 & 0.54 $\pm$ 0.06 & 0.55 $\pm$ 0.11 & \xmark \\
 & CIFAR-10 & Deer $\rightarrow$ Horse & 0.90 $\pm$ 0.01 & 0.89 $\pm$ 0.01 & 0.86 $\pm$ 0.03 & \xmark \\
 & CIFAR-10 & Ship $\rightarrow$ Truck & 0.93 $\pm$ 0.02 & 0.92 $\pm$ 0.02 & 0.85 $\pm$ 0.02 & \xmark \\
 & DermaMNIST & NV $\rightarrow$ Vasc & 0.94 $\pm$ 0.02 & 0.94 $\pm$ 0.02 & 0.94 $\pm$ 0.02 & \xmark \\
 & DermaMNIST & Mel $\rightarrow$ DF & 0.94 $\pm$ 0.01 & 0.97 $\pm$ 0.01 & 0.96 $\pm$ 0.01 & \xmark \\
 & PathMNIST & BG $\rightarrow$ Deb & 0.74 $\pm$ 0.12 & 0.75 $\pm$ 0.11 & 0.79 $\pm$ 0.12 & \xmark \\
 & PathMNIST & LymP $\rightarrow$ Deb & 0.99 $\pm$ 0.01 & 0.35 $\pm$ 0.04 & 0.68 $\pm$ 0.08 & \xmark \\
\midrule
\textbf{ViT} & BloodMNIST & Baso $\rightarrow$ Eos & 0.75 $\pm$ 0.02 & 0.92 $\pm$ 0.00 & 0.93 $\pm$ 0.02 & \xmark \\
 & BloodMNIST & LymB $\rightarrow$ Mono & 0.44 $\pm$ 0.12 & 0.78 $\pm$ 0.07 & 0.71 $\pm$ 0.08 & \xmark \\
 & CIFAR-10 & Auto $\rightarrow$ Truck & 0.95 $\pm$ 0.06 & 0.95 $\pm$ 0.05 & 0.93 $\pm$ 0.03 & \xmark \\
 & CIFAR-10 & Truck $\rightarrow$ Auto & 0.89 $\pm$ 0.08 & 0.83 $\pm$ 0.08 & 0.84 $\pm$ 0.07 & \xmark \\
 & DermaMNIST & BCC $\rightarrow$ Mel & 0.99 $\pm$ 0.00 & 0.81 $\pm$ 0.14 & 0.74 $\pm$ 0.11 & \xmark \\
 & DermaMNIST & Mel $\rightarrow$ DF & 0.99 $\pm$ 0.00 & 0.84 $\pm$ 0.07 & 0.88 $\pm$ 0.03 & \xmark \\
 & PathMNIST & Adip $\rightarrow$ BG & 1.00 & 1.00 $\pm$ 0.00 & 1.00 & \xmark \\
 & PathMNIST & LymP $\rightarrow$ Deb & 0.96 $\pm$ 0.00 & 0.87 $\pm$ 0.01 & 0.79 $\pm$ 0.03 & \xmark \\
\bottomrule
\end{tabular}
}
\end{table*}

\section{Discussion and conclusion}
\label{sec:discussion_conclusion}
This work demonstrates an extension of the method introduced by~\citet{goldwasser2024ieee, goldwasser2024arxiv}, providing a formal guarantee of cryptographic undetectability for backdoors planted via spiked covariance, from simplified neural networks to modern transformer and convolutional architectures.
Given the generality of our approach and the universal presence of CAVs in neural representations, no theoretical restrictions prevent extension to other architectures. 
As part of our analysis, we provide theoretical justification for the impracticability of distinguishing a backdoored model from a clean version, given the computational hardness of the hypothesis test discussed in Sec.~\ref{sec:backdoor_sota_networks}.
Future work should aim to derive a formal proof to establish cryptographic undetectability. Such a proof would likely require an analytical characterisation of trained neural network weight distributions, which remains an open problem.
The practical utility of such a theoretical result is arguably limited, given that parameter-space defences are ineffective against our backdoor in experiments.
Indeed, prior work typically establishes undetectability through empirical evaluation against such defences alone; our analysis meets this standard and additionally provides a formal proof in the idealised setting.

A key aspect of our approach is that any undetectability is baked into the weights themselves; the input trigger is secondary, derived via optimisation once the backdoor is planted. This decoupling allows users to explicitly control visual imperceptibility according to their specific requirements without compromising the undetectability of the backdooring mechanism itself. Depending on the data domain, further refinement of the optimisation process can yield more sophisticated, human-imperceptible triggers.

\paragraph{Limitations} Our work lacks a proof of full white-box undetectability. While we provide theoretical justification and empirical evidence, we cannot conclude with theoretical guarantees. The undetectability claim is conjectural and restricted to the stated MLaaS threat model. Moreover, the difficulty of the proposed hypothesis test depends on the hyperparameter value $\theta$. At present, we lack a well-defined range for selecting this parameter, beyond the general intuition that $\theta$ should be small to render the test harder to falsify. The same challenge applies to the dimensions $d$ and $m$.
Additionally, although the experiments are limited to the domain of computer vision, the approach is extendable to other domains such as natural language processing. Future work may also involve evaluations on other datasets and comparisons with existing backdooring approaches.

\bibliographystyle{plainnat} 
\bibliography{references}

\clearpage

\appendix
\FloatBarrier
\section{Visualisation of spiked covariance}
\label{app:visualisation_spiked_covariance}

\begin{figure}[H]
    \centering

    \begin{subfigure}{0.3\textwidth}
        \centering
        \includegraphics[width=\linewidth]{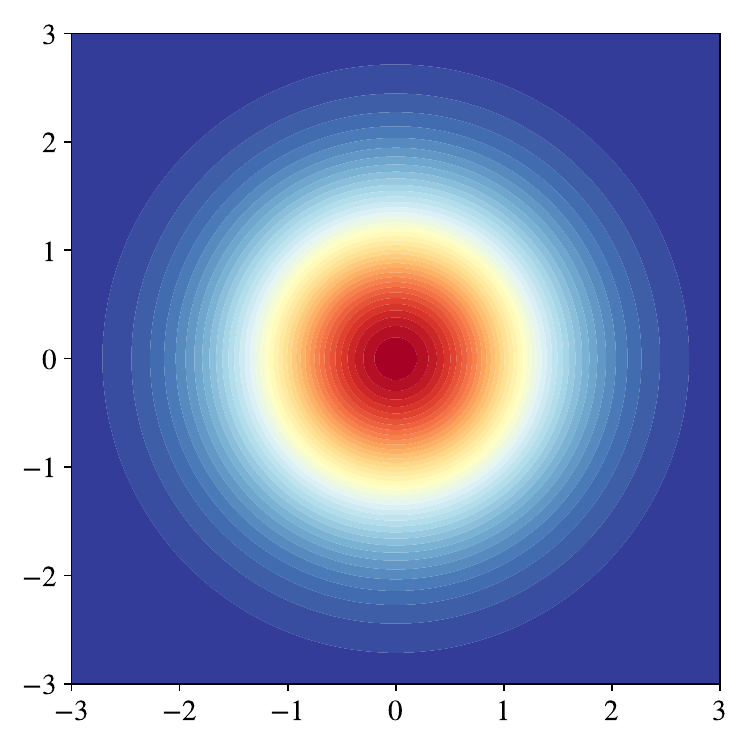}
        \caption{}
    \end{subfigure}
    \hspace{0.02\textwidth}
    \begin{subfigure}{0.3\textwidth}
        \centering
        \includegraphics[width=\linewidth]{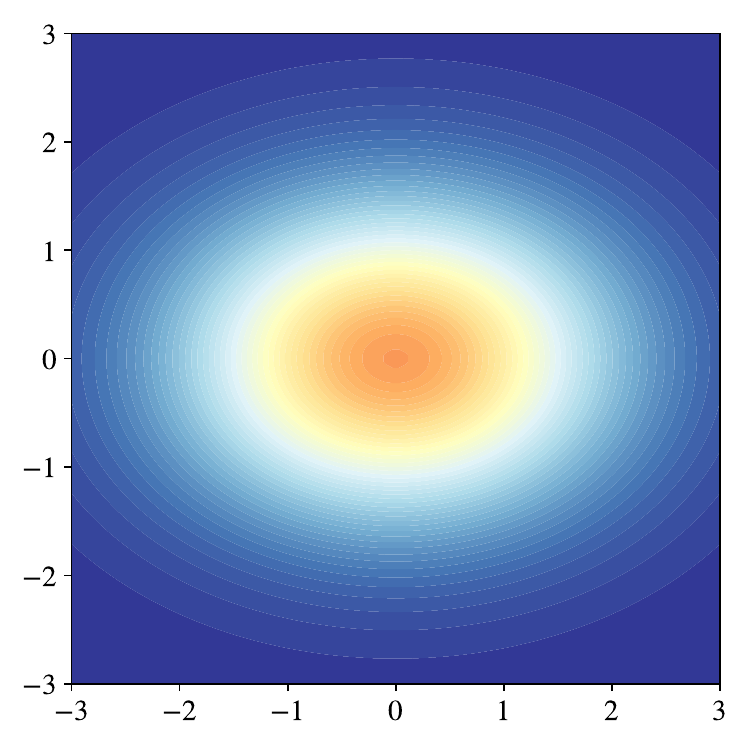}
        \caption{}
    \end{subfigure}

    \caption{(a) The clean distribution follows a standard multivariate Gaussian, which is modified with a spiked covariance to obtain (b) the backdoored distribution. Both are shown as 2D contour lines using a shared colour scale.}
    \label{fig:covariance_spike}
\end{figure}

Figure \ref{fig:covariance_spike} illustrates the difference between clean and backdoored distributions, highlighting the increased variance introduced by the spiked covariance. The clean distribution follows a standard multivariate Gaussian $\mathcal{N}(\mathbf{0}, I_d)$, while the backdoored distribution $\mathcal{N}(\mathbf{0}, I_d + \theta \boldsymbol{\nu} \boldsymbol{\nu}^\top)$ incorporates a low-rank covariance perturbation, creating a spike along a secret direction $\boldsymbol{\nu} \in \mathbb{R}^d$. While statistically indistinguishable under certain conditions, the variance is visibly higher in the latter for a high value $\theta$. Refer to Sec.~\ref{sec:backdoor_random_relu_networks} for further details. 

\FloatBarrier
\section{Spiked covariance distribution}
\label{app:spiked_covariance_distribution}

The following shows that the weight transformation from $W^{(\ell_{\text{bd}})}$ to $\hat{W}^{(\ell_{\text{bd}})}$, both defined in Sec.\ref{sec:backdoor_sota_networks}, would be statistically identical to sampling new weights from a distribution $\mathbf{g}_i \sim \mathcal{N}(\mathbf{0}, \sigma^2 I_m + \theta \boldsymbol{\nu}_\text{bd} \boldsymbol{\nu}_\text{bd}^\top)$, under the assumption that $W^{(\ell_{\text{bd}})}$ was Gaussian.

Assume each column of the weight matrix $W^{(\ell_{\text{bd}})}$, denoted $\mathbf{w} \in \mathbb{R}^m$, are sampled from a distribution $\mathbf{w} \sim \mathcal{N}(\mathbf{0}, \sigma^2 I_m)$, which gives covariance $\text{Cov}(\mathbf{w}) = \mathbb{E}[\mathbf{w}\mathbf{w}^\top] = \sigma^2 I_m$.\footnote{This assumes the weight vector $\mathbf{w}$ is i.i.d (independent and identically distributed) with zero mean, a shared variance $\sigma^2$, and zero pairwise covariance ($\mathbb{E}[w_i, w_j] = 0, \quad \forall i \neq j$ ).} The covariance of these weights scaled by the standard deviation $\sigma$ is,
\begin{equation}
    \text{Cov}\left( \frac{\mathbf{w}}{\sigma} \right) = \left( \frac{1}{\sigma} \right) (\sigma^2 I_m) \left( \frac{1}{\sigma} \right)^\top = \frac{\sigma^2}{\sigma^2} I_m = I_m. \footnote{Covariance property for a vector $\mathbf{x}$ and deterministic matrix $A$: $\text{Cov}(A\mathbf{x}) = A\text{Cov}(\mathbf{x})A^\top$.} 
    \label{eq:cov_w_sigma}
\end{equation} 
The identity covariance makes the scaled weights mathematically indistinguishable from a sample of the standard Gaussian, as its mean is assumed to be zero. 

The covariance of the backdoored weights $\hat{W}^{(\ell_{\text{bd}})}$ when applying the Cholesky factor $L$ as in Sec.~\ref{sec:backdoor_sota_networks} is, when using the result from Eq.~\eqref{eq:cov_w_sigma}, 
\begin{equation}
\begin{aligned}
\mathrm{Cov}(\hat{\mathbf{w}})
&= \mathrm{Cov}\!\left(L \frac{\mathbf{w}}{\sigma}\right)
 = L \, \mathrm{Cov}\!\left(\frac{\mathbf{w}}{\sigma}\right) L^{T} \\
&= L I_d L^{T}
 = L L^{T}
 = \Sigma,
\end{aligned}
\end{equation}
where $\Sigma$ is defined in Eq.~\eqref{eq:spike_covariance_sigma}.

As discussed in Sec.~\ref{sec:backdoor_sota_networks}, the weights do not strictly follow a Gaussian distribution, but the above argument motivates our approach.

\newpage
\section{Empirical structure of the weight covariance}
\label{app:structure_weight_covariance}
In conflict with our simplified assumption, the covariance of the model weights is not diagonal; model training induces correlations and the weights are tied together via the loss landscape. Still, when viewed through the spectrum of the Hessian (the second derivatives of the loss with respect to the weights), the bulk behaves as if the covariance were essentially isotropic and degenerate, with a small number of outliers carrying the structured signal. This was shown empirically by \citet{sagun2017empirical, sagun2018empirical}, who decomposed the Hessian via the Generalised Gauss-Newton form into a covariance-of-gradients term plus a residual.

Both at initialisation and at the end of training, almost all Hessian eigenvalues sit in a narrow bulk concentrated near zero, with a handful of large positive outliers detached from it \citep{sagun2016eigenvalues, sagun2017empirical, sagun2018empirical}. The two parts have different origins: the bulk is governed by the architecture, while the outliers are governed by the data. \citet{sagun2017empirical, sagun2018empirical} show empirically that for a $k$-class classification problem, the number of outliers above the bulk matches $k$, which is exactly what one would expect if the dominant non-trivial directions come from the rank-$k$ structure of the gradient-of-outputs covariance in their decomposition.

Increasing layer width does not create new outlier directions; it only inflates the bulk near zero \citep{sagun2016eigenvalues, sagun2017empirical, sagun2018empirical}. Briefly put: most directions in overparameterised networks correspond to redundant zero modes, alongside whatever sparse structured signal the data contributes.
Importantly for the present context, this low effective rank is visible directly in the weight matrices, not only the Hessian. The parallel observation was made by \citet{martin2021implicit}, who find that trained weight matrices develop strongly non-random, often heavy-tailed spectra, with structure aligned to the data. They find that, over the course of training, the information in the weight matrices increasingly concentrates in a sparse spike, and that there is likely no simple low rank approximation for the weight matrices.

Despite the important off-diagonal directions, the empirical results in Sec.~\ref{sec:results_analysis} demonstrate that the model's learned behaviour is preserved when using the spiked covariance in Eq.~\ref{eq:spike_covariance_sigma} to replace weights in the backdoor layer. 

\FloatBarrier
\newpage
\section{Optimisation algorithm for input space trigger}
\label{app:optimisation_input_trigger}

Algorithm \ref{alg:optimisation_input_trigger} outlines the optimisation procedure for obtaining the input trigger using the backdoored model denoted $\hat{f}$. In particular, we define $\mathbf{z}_\text{source}^{(\ell_{\text{bd}})} \in \mathbb{R}^{m}$ as the activation of an input from the source class in layer $\ell_{\text{bd}}$, and $\hat{\mathbf{z}}_\text{source}^{(\ell_{\text{bd}})}$ as the activation from the corresponding triggered input. The main optimisation objective is to align the latter with the spike direction $\boldsymbol{\nu}_{\text{bd}} \in \mathbb{R}^m$. 

Stealthiness requires perturbations to remain sufficiently subtle such that a triggered input $\hat{\mathbf{x}}$ is difficult to distinguish from its clean counterpart $\mathbf{x}$. Ideally, this similarity should hold for both human observers and automated anomaly detection systems. The optimisation accounts for stealthiness by including L2 regularisation to its objective, constrain pixel values, and enforce sparsity via a threshold $\tau$. The optimisation is carried out on a smaller patch $\mathbb{R}^{h_s \times w_s \times c}$, which is subsequently upsampled to induce a smoothing (blurring) effect.

\begin{algorithm} [htbp]
\caption{Optimise trigger perturbation for input space}
\begin{algorithmic}[1]
\State \textbf{Input:} $\mathbf{x}$, $\boldsymbol{\nu}_{\text{bd}}$, $\hat{f}$, steps, lr, scale, $\lambda_{\text{L2}}$, $\epsilon$, $\tau$
\State \textbf{Output:} $\boldsymbol{\nu}_{\text{in}}$

\State $(b, c, h, w) \gets \text{shape}(\mathbf{x})$
\State $h_s \gets \lfloor h / \text{scale} \rfloor$
\State $w_s \gets \lfloor w / \text{scale} \rfloor$

\State $\mathbf{\delta} \gets \mathbf{0}^{1 \times c \times h_s \times w_s}$
\State $\mathbf{\delta}.\text{requires\_grad} \gets \text{True}$

\State Initialize optimizer $\gets$ Adam($\mathbf{\delta}, \text{lr}$)
\State $\boldsymbol{\nu}_{\text{bd}} \gets \text{normalize}(\boldsymbol{\nu}_{\text{bd}})$
\State $\text{dataloader} \gets \text{get\_dataloader}(\mathbf{x})$

\For{$\text{step} = 1$ to $\text{steps}$}
    \State $\text{optimizer.zero\_grad}()$

    \For{each batch $\mathbf{x}_{\text{batch}}$ in dataloader}
        
        \State $\boldsymbol{\nu}_{\text{in}} \gets \text{interpolate}(\mathbf{\delta}, (h, w))$
        \State $\boldsymbol{\nu}_{\text{in}}' \gets \text{clamp}(\boldsymbol{\nu}_{\text{in}}, -\epsilon, \epsilon)$
        
        \State $\mathbf{y} \gets \hat{f}(\mathbf{x}_{\text{batch}} + \boldsymbol{\nu}_{\text{in}}')$
        \State $\mathbf{h} \gets \text{activation}$

        \State $\text{alignment} \gets - \frac{1}{|\mathbf{x}_{\text{batch}}|} \sum (\mathbf{h} \cdot \boldsymbol{\nu}_{\text{bd}})$
        
        \State $\text{L2} \gets \frac{\lambda_{\text{L2}}}{|\mathbf{x}_{\text{batch}}|} \sum (\boldsymbol{\nu}_{\text{in}}')^2$

        \State $\text{loss} \gets \text{alignment} + \text{L2}$

        \State $\text{loss.backward}()$
        \State $\text{activation} \gets \emptyset$
    \EndFor

    \State $\text{optimizer.step}()$

    \State $\mathbf{\delta} \gets \text{sign}(\mathbf{\delta}) \cdot \max(|\mathbf{\delta}| - \tau, 0)$

\EndFor

\State $\boldsymbol{\nu}_{\text{in}} \gets \text{interpolate}(\mathbf{\delta}, (h, w))$
\State \Return $\text{squeeze}(\boldsymbol{\nu}_{\text{in}})$

\end{algorithmic}
\label{alg:optimisation_input_trigger}
\end{algorithm}

\FloatBarrier
\newpage
\section{Notation and symbols reference}
\label{app:symbols_and_notation}

Table~\ref{tab:symbols_and_notation} summarises the main symbols and notations used throughout the paper, most of which are introduced in Sec.\ref{sec:theoretical_background_and_problem_formulation}.

{
\setlength{\tabcolsep}{4pt}
\begin{xltabular}{\linewidth}{
  @{}
  >{\raggedright\arraybackslash}p{1.9cm}
  l
  >{\raggedright\arraybackslash}X
  >{\raggedright\arraybackslash}p{1.6cm}
  @{}
}
\caption{Notation used throughout the paper. The reference column indicates where each symbol is introduced or used.}
\label{tab:notation}\\
\toprule
\textbf{Category} & \multicolumn{1}{l}{\textbf{Symbol}} &
\textbf{Description} & \textbf{Reference} \\
\midrule
\endfirsthead

\multicolumn{4}{@{}l}{\textit{Table~\thetable\ continued}}\\
\toprule
\textbf{Category} & \multicolumn{1}{l}{\textbf{Symbol}} &
\textbf{Description} & \textbf{Reference} \\
\midrule
\endhead

\midrule
\multicolumn{4}{r@{}}{\textit{continued on next page}}\\
\endfoot

\bottomrule
\endlastfoot

Networks
  & $f$ & Clean classifier
  & Sec.~\ref{sec:backdoor_sota_networks} \\
  & $\hat f$ & Backdoored classifier
  & Sec.~\ref{sec:backdoor_sota_networks} \\
  & $\ell_{\mathrm{bd}}$ & Backdoor layer: an additional linear layer inserted before the output head
  & Sec.~\ref{sec:backdoor_sota_networks} \\

\midrule

Dimensions
  & $h,\ w,\ c$ & Input image height, width and number of channels
  & Sec.~\ref{sec:backdoor_sota_networks} \\
  & $k$ & Number of output classes
  & Sec.~\ref{sec:backdoor_sota_networks} \\
  & $d$ & Input dimension of layer $\ell_{\mathrm{bd}}$
  & Sec.~\ref{sec:backdoor_random_relu_networks} \\
  & $m$ & Output dimension of layer $\ell_{\mathrm{bd}}$
  & Sec.~\ref{sec:backdoor_random_relu_networks} \\

\midrule
Data
  & $\mathbf{x}$ & Clean input
  & Sec.~\ref{sec:backdoor_random_relu_networks} \\
  & $\hat{\mathbf{x}}$ & Triggered input
  & Sec.~\ref{sec:backdoor_random_relu_networks} \\
  & $\mathbf{x}_{\text{source}}$,\ $\mathbf{x}_{\text{target}}$
  & Inputs from the source and target classes
  & Sec.~\ref{sec:backdoor_sota_networks} \\
  & $\mathcal{D}_{\text{source}}$,\ $\mathcal{D}_{\text{target}}$
  & Equally sized sets of source- and target-class inputs used to fit a logistic regression probe
  & Sec.~\ref{sec:backdoor_sota_networks} \\

\midrule

\smash{\begin{tabular}[t]{@{}l@{}}
Weights and\\
covariance
\end{tabular}}
  & $W^{(\ell_{\mathrm{bd}})}$ & Learned weights of the backdoor layer (clean model)
  & Sec.~\ref{sec:backdoor_sota_networks} \\
  & $\hat{W}^{(\ell_{\mathrm{bd}})}$ & Backdoored weights
  & Eq.~\eqref{eq:w_bd} \\
  & $\sigma^2$ & Scalar variance of the learned weights in layer $\ell_{\mathrm{bd}}$
  & Sec.~\ref{sec:backdoor_sota_networks} \\
  & $\theta$ & Spike strength
  & Eq.~\eqref{eq:spike_covariance}--\eqref{eq:spike_covariance_sigma} \\
  & $\Sigma$ & Spiked covariance
  & Eq.~\eqref{eq:spike_covariance_sigma} \\
  & $L$ & Lower-triangular Cholesky factor, $LL^\top=\Sigma$
  & Sec.~\ref{sec:backdoor_sota_networks} \\
  & $I_d,\ I_m$ & Identity matrices
  & Eq.~\eqref{eq:spike_covariance}--\eqref{eq:spike_covariance_sigma} \\

\midrule
\smash{\begin{tabular}[t]{@{}l@{}}
Secret keys\\
and trigger
\end{tabular}}
  & $\boldsymbol{\nu}_{\mathrm{bd}}$ & Secret key in the activation space of $\ell_{\mathrm{bd}}$: sparsified, unit-norm normal vector of the source/target logistic regression boundary
  & Eq.~\eqref{eq:spike_covariance_sigma} \\
  & $\boldsymbol{\nu}_{\mathrm{in}}$ & Secret key in input space, optimised such that the triggered activations align with $\boldsymbol{\nu}_{\mathrm{bd}}$
  & Sec.~\ref{sec:backdoor_sota_networks} \\
  & $\lambda$ & Trigger weight
  & Sec.~\ref{sec:backdoor_random_relu_networks} \\

\midrule
Trigger optimisation
  & $\lambda_{\text{L2}}$ & Weight of the L2 penalty
  & Sec.~\ref{sec:backdoor_sota_networks} 
  
\label{tab:symbols_and_notation}
\end{xltabular}
}

\FloatBarrier
\section{The Marchenko-Pastur theorem}
\label{app:marchenko-pastur}

The Marchenko-Pastur theorem is a fundamental result in random matrix theory that describes the eigenvalues of large sample covariance matrices. Assume a data matrix $X \in \mathbb{R}^{p \times n}$, where $p$ denotes the number of variables and $n$ the number of samples, in which entries are independent identically distributed random variables with mean $0$ and variance $\sigma^2$. The sample covariance matrix is expressed as
\begin{equation}
    S = \frac{1}{n}XX^\top \in \mathbb{R}^{p \times p}.
\end{equation}

When the population covariance is $\sigma^2 I$, the eigenvalues of $S$ concentrate on the interval $[\sigma^2(1 - \sqrt{\gamma})^2, \sigma^2(1 + \sqrt{\gamma})^2]$ when $p, n \rightarrow \infty$ with $\frac{p}{n} \rightarrow \gamma \in (0, +\infty)$. This shows that even in the absence of structure, the eigenvalues spread over a non-trivial interval purely due to noise. Furthermore, spikes or dominant principal directions are only distinguishable from random noise if their associated eigenvalues exceed the upper bound. 

\newpage
\section{Hyperparameter configurations}
\label{app:hyperparameter_config}

This section summarises the hyperparameter configurations used in the experiments reported in Sec.~\ref{sec:results_analysis}. All experiments -- across both architectures and datasets -- share the common settings listed in Tab.~\ref{tab:hyperparameter_config_shared}. A fixed scaling factor $\lambda = 0.5$ is used for the input trigger in ResNet18 experiments, whereas a value of $\lambda = 2$ is used for ViT experiments. The only hyperparameter varying across source-target class pairs is the L2 regularisation coefficient $\lambda_\text{L2}$ in the optimisation procedure (Alg.~\ref{alg:optimisation_input_trigger}). These values are reported in Tab.~\ref{tab:hyperparameter_config_l2}.

\begin{table}[htbp]
\centering
\caption{Shared hyperparameter values across architectures and datasets.}
\label{tab:hyperparameter_config_shared}
\begin{tabular}{l l c}
\toprule
& \textbf{Hyperparameter} & \textbf{Value} \\
\midrule
\textbf{Optimisiation of input trigger} & steps & 200   \\
 & scale & 4 \\
& lr & 0.01 \\
 & $\epsilon$ &  30/255 \\
 & $\tau$ & 0.001 \\
\textbf{Sparsity} & $\alpha$ & 0.1 \\
\textbf{Strength of spiked covariance} & $\theta$ & 0.1 \\
\bottomrule
\end{tabular}
\end{table}

\begin{table}[htbp]
\caption{The hyperparameter value of the L2 regularisation coefficient $\lambda_{\text{L2}}$ in Alg.~\ref{alg:optimisation_input_trigger} for all source-target class pairs across datasets. The full class names corresponding to the abbreviations are listed in App.~\ref{app:label_abbreviations}.}
\label{tab:hyperparameter_config_l2}
\centering
\begin{tabular}{l l l c}
\toprule
\textbf{Model} & \textbf{Dataset} & \textbf{Source $\rightarrow$ Target} & \textbf{Value} $\lambda_{\text{L2}}$   \\
\midrule
\textbf{ResNet18} & BloodMNIST & Baso $\rightarrow$ Eos & 15 \\
 & BloodMNIST & EryBl $\rightarrow$ Baso & 6.5 \\
 & CIFAR-10 & Deer $\rightarrow$ Horse & 1.5  \\
 & CIFAR-10 & Ship $\rightarrow$ Truck & 3  \\
 & DermaMNIST & NV $\rightarrow$ Vasc & 10  \\
 & DermaMNIST & Mel $\rightarrow$ DF & 2.5 \\
 & PathMNIST & BG $\rightarrow$ Deb & 8  \\
 & PathMNIST & LymP $\rightarrow$ Deb & 4  \\
\midrule
\textbf{ViT} & BloodMNIST & Baso $\rightarrow$ Eos & 16.5  \\
 & BloodMNIST & LymB $\rightarrow$ Mono & 5  \\
 & CIFAR-10 & Auto $\rightarrow$ Truck & 1 \\
 & CIFAR-10 & Truck $\rightarrow$ Auto & 1  \\
 & DermaMNIST & BCC $\rightarrow$ Mel & 5.5 \\
 & DermaMNIST & Mel $\rightarrow$ DF & 8  \\
 & PathMNIST & Adip $\rightarrow$ BG & 1 \\
 & PathMNIST & LymP $\rightarrow$ Deb & 1 \\
\bottomrule
\end{tabular}
\end{table}

\FloatBarrier

\section{Experiments compute resources} 
\label{app:compute_resources}
All computations were performed on an HPC cluster using NVIDIA V100 or A100 GPUs. Comparable hardware is not strictly required; rather, sufficient GPU memory to accommodate the model size and data is the primary requirement. 

\newpage

\section{Class label abbreviations}
\label{app:label_abbreviations}

Table \ref{tab:label_abbreviations} provides abbreviations of class names in all evaluated datasets BloodMNIST, CIFAR-10, DermaMNIST and PathMNIST.

\begin{table}[H]
\caption{Class label abbreviations}
\centering
\begin{tabular}{ll}
\hline
\textbf{Full name} & \textbf{Abbreviation} \\
\hline

Basophil & Baso \\
Eosinophil & Eos \\
Erythroblast & EryBl \\
Immature granulocytes & ImmGran \\
Lymphocyte & LymB \\
Monocyte & Mono \\
Neutrophil & Neut \\
Platelet & Plt \\

\\[0.6em]
Airplane & Air \\
Automobile & Auto \\
Bird &  Bird \\
Cat & Cat \\
Deer & Deer \\
Dog & Dog \\
Frog & Frog \\
Horse & Horse \\
Ship & Ship \\
Truck & Truck \\

\\[0.6em]
Actinic keratoses and intraepithelial carcinoma & AKIEC \\
Basal cell carcinoma & BCC \\
Benign keratosis-like lesions & BKL \\
Dermatofibroma & DF \\
Melanoma & Mel \\
Melanocytic nevi & NV \\
Vascular lesions & Vasc \\

\\[0.6em]
Adipose & Adip \\
Background & BG \\
Cancer-associated stroma & CAS \\
Colorectal adenocarcinoma epithelium & CAE \\
Debris & Deb \\
Lymphocytes & LymP \\
Mucus & Muc \\
Normal colon mucosa & NCM \\
Smooth muscle & SM \\

\hline

\end{tabular}
\label{tab:label_abbreviations}
\end{table}

\newpage
\section{Performance comparison of defences on clean vs. triggered inputs}
\label{app:additional_results_defences}

Table \ref{tab:additional_results_defences} compares the performance of defence methods pruning, parameter clipping and parameter noise injection on both clean and triggered inputs. Notably, the triggered results match those in Tab.~\ref{tab:results_defences}, while this table additionally reports performance on clean inputs for comparison.

\begin{table}[H]
\caption{Performance comparison of defence methods pruning, parameter clipping and parameter noise injection on clean and triggered inputs for ResNet18 and ViT across four datasets. \textit{Clean} denotes classification accuracy on unmodified inputs from all classes, while \textit{triggered} indicates the ASR on triggered source-target class pairs. Uncertainty is reported as lower and upper bounds of the 95\% confidence interval, with all values rounded to two decimal places. All hyperparameter configurations are provided in App.~\ref{app:hyperparameter_config}, with full class names corresponding to the abbreviations listed in App.~\ref{app:label_abbreviations}.}
\label{tab:additional_results_defences}
\centering
{\tiny
\begin{tabular}{l l l c c c c c c}
\toprule
 & & & \multicolumn{2}{c}{\textbf{Pruning}} & \multicolumn{2}{c}{\textbf{Clipping}} & \multicolumn{2}{c}{\textbf{Noise}} \\
\cmidrule(lr){4-5} \cmidrule(lr){6-7} \cmidrule(lr){8-9}
\textbf{Model} & \textbf{Dataset} & \textbf{Source $\rightarrow$ Target} & \textbf{Clean} & \textbf{Triggered} & \textbf{Clean} & \textbf{Triggered} & \textbf{Clean} & \textbf{Triggered} \\
\midrule
\textbf{ResNet18} & BloodMNIST & Baso $\rightarrow$ Eos & 0.93 $\pm$ 0.00 & 1.00 $\pm$ 0.00 & 0.94 $\pm$ 0.00 & 0.73 $\pm$ 0.05 & 0.95 $\pm$ 0.00 & 0.89 $\pm$ 0.05 \\
 & BloodMNIST & EryBl $\rightarrow$ Baso & 0.92 $\pm$ 0.00 & 0.66 $\pm$ 0.08 & 0.96 $\pm$ 0.00 & 0.54 $\pm$ 0.06 & 0.94 $\pm$ 0.00 & 0.55 $\pm$ 0.11 \\
 & CIFAR-10 & Deer $\rightarrow$ Horse & 0.84 $\pm$ 0.00 & 0.90 $\pm$ 0.01 & 0.88 $\pm$ 0.00 & 0.89 $\pm$ 0.01 & 0.87 $\pm$ 0.00 & 0.86 $\pm$ 0.03 \\
 & CIFAR-10 & Ship $\rightarrow$ Truck & 0.87 $\pm$ 0.01 & 0.93 $\pm$ 0.02 & 0.90 $\pm$ 0.01 & 0.92 $\pm$ 0.02 & 0.88 $\pm$ 0.00 & 0.85 $\pm$ 0.02 \\
 & DermaMNIST & NV $\rightarrow$ Vasc & 0.65 $\pm$ 0.01 & 0.94 $\pm$ 0.02 & 0.66 $\pm$ 0.01 & 0.94 $\pm$ 0.02 & 0.67 $\pm$ 0.01 & 0.94 $\pm$ 0.02 \\
 & DermaMNIST & Mel $\rightarrow$ DF & 0.62 $\pm$ 0.01 & 0.94 $\pm$ 0.01 & 0.70 $\pm$ 0.00 & 0.97 $\pm$ 0.01 & 0.69 $\pm$ 0.01 & 0.96 $\pm$ 0.01 \\
 & PathMNIST & BG $\rightarrow$ Deb & 0.78 $\pm$ 0.02 & 0.74 $\pm$ 0.12 & 0.79 $\pm$ 0.02 & 0.75 $\pm$ 0.11 & 0.80 $\pm$ 0.01 & 0.79 $\pm$ 0.12 \\
 & PathMNIST & LymP $\rightarrow$ Deb & 0.80 $\pm$ 0.01 & 0.99 $\pm$ 0.01 & 0.84 $\pm$ 0.00 & 0.35 $\pm$ 0.04 & 0.81 $\pm$ 0.01 & 0.68 $\pm$ 0.08 \\
\midrule
\textbf{ViT} & BloodMNIST & Baso $\rightarrow$ Eos & 0.96 $\pm$ 0.00 & 0.75 $\pm$ 0.02 & 0.96 $\pm$ 0.00 & 0.92 $\pm$ 0.00 & 0.95 $\pm$ 0.01 & 0.93 $\pm$ 0.02 \\
 & BloodMNIST & LymB $\rightarrow$ Mono & 0.95 $\pm$ 0.00 & 0.44 $\pm$ 0.12 & 0.98 $\pm$ 0.00 & 0.78 $\pm$ 0.07 & 0.96 $\pm$ 0.00 & 0.71 $\pm$ 0.08 \\
 & CIFAR-10 & Auto $\rightarrow$ Truck & 0.91 $\pm$ 0.00 & 0.95 $\pm$ 0.06 & 0.93 $\pm$ 0.00 & 0.95 $\pm$ 0.05 & 0.91 $\pm$ 0.00 & 0.93 $\pm$ 0.03 \\
 & CIFAR-10 & Truck $\rightarrow$ Auto & 0.91 $\pm$ 0.00 & 0.89 $\pm$ 0.08 & 0.93 $\pm$ 0.00 & 0.83 $\pm$ 0.08 & 0.91 $\pm$ 0.00 & 0.84 $\pm$ 0.07 \\
 & DermaMNIST & BCC $\rightarrow$ Mel & 0.75 $\pm$ 0.02 & 0.99 $\pm$ 0.00 & 0.72 $\pm$ 0.03 & 0.81 $\pm$ 0.14 & 0.69 $\pm$ 0.02 & 0.74 $\pm$ 0.11 \\
 & DermaMNIST & Mel $\rightarrow$ DF & 0.74 $\pm$ 0.02 & 0.99 $\pm$ 0.00 & 0.67 $\pm$ 0.03 & 0.84 $\pm$ 0.07 & 0.69 $\pm$ 0.01 & 0.88 $\pm$ 0.03 \\
 & PathMNIST & Adip $\rightarrow$ BG & 0.86 $\pm$ 0.00 & 1.00 & 0.94 $\pm$ 0.01 & 1.00 $\pm$ 0.00 & 0.90 $\pm$ 0.00 & 1.00 \\
 & PathMNIST & LymP $\rightarrow$ Deb & 0.87 $\pm$ 0.00 & 0.96 $\pm$ 0.00 & 0.90 $\pm$ 0.00 & 0.87 $\pm$ 0.01 & 0.88 $\pm$ 0.00 & 0.79 $\pm$ 0.03 \\
\bottomrule
\end{tabular}
}
\end{table}

\FloatBarrier
\newpage

\section{Performance of fine-tuning and fine-pruning defences}
\label{app:additional_results_tuning_defence}

\begin{figure}[h]
\centering

\begin{subfigure}{0.5\textwidth}
    \includegraphics[width=\textwidth]{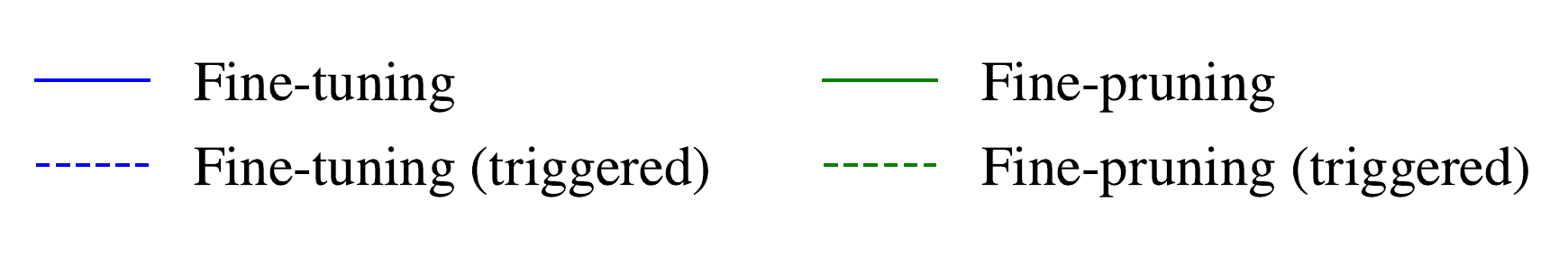}
\end{subfigure}

\vspace{0.2cm}

\begin{subfigure}{0.185\textwidth}
    \includegraphics[width=\linewidth]{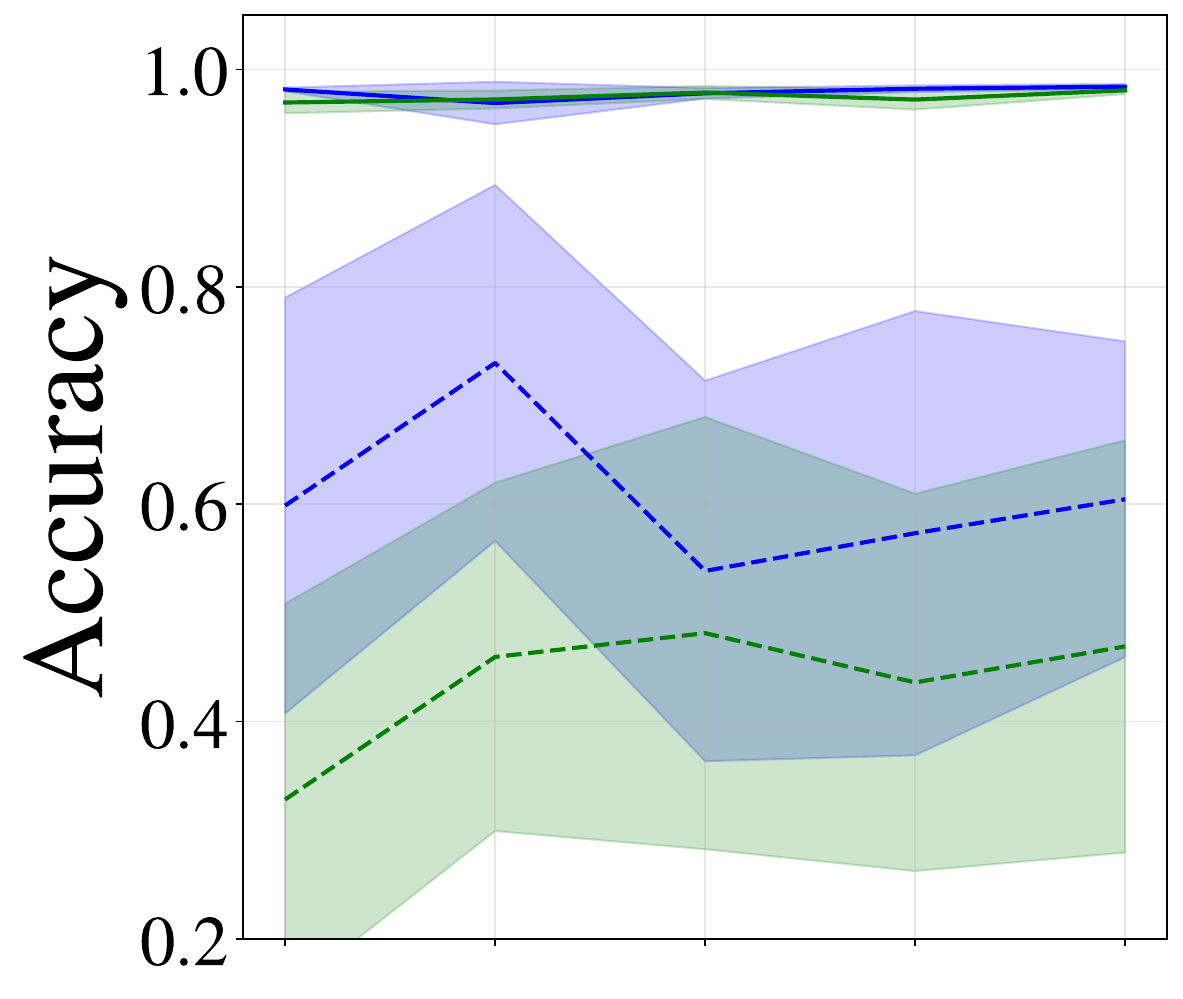}
    \caption{EryBl-Baso}
\end{subfigure} 
\hspace{0.02\textwidth}
\begin{subfigure}{0.185\textwidth}
    \includegraphics[width=\linewidth]{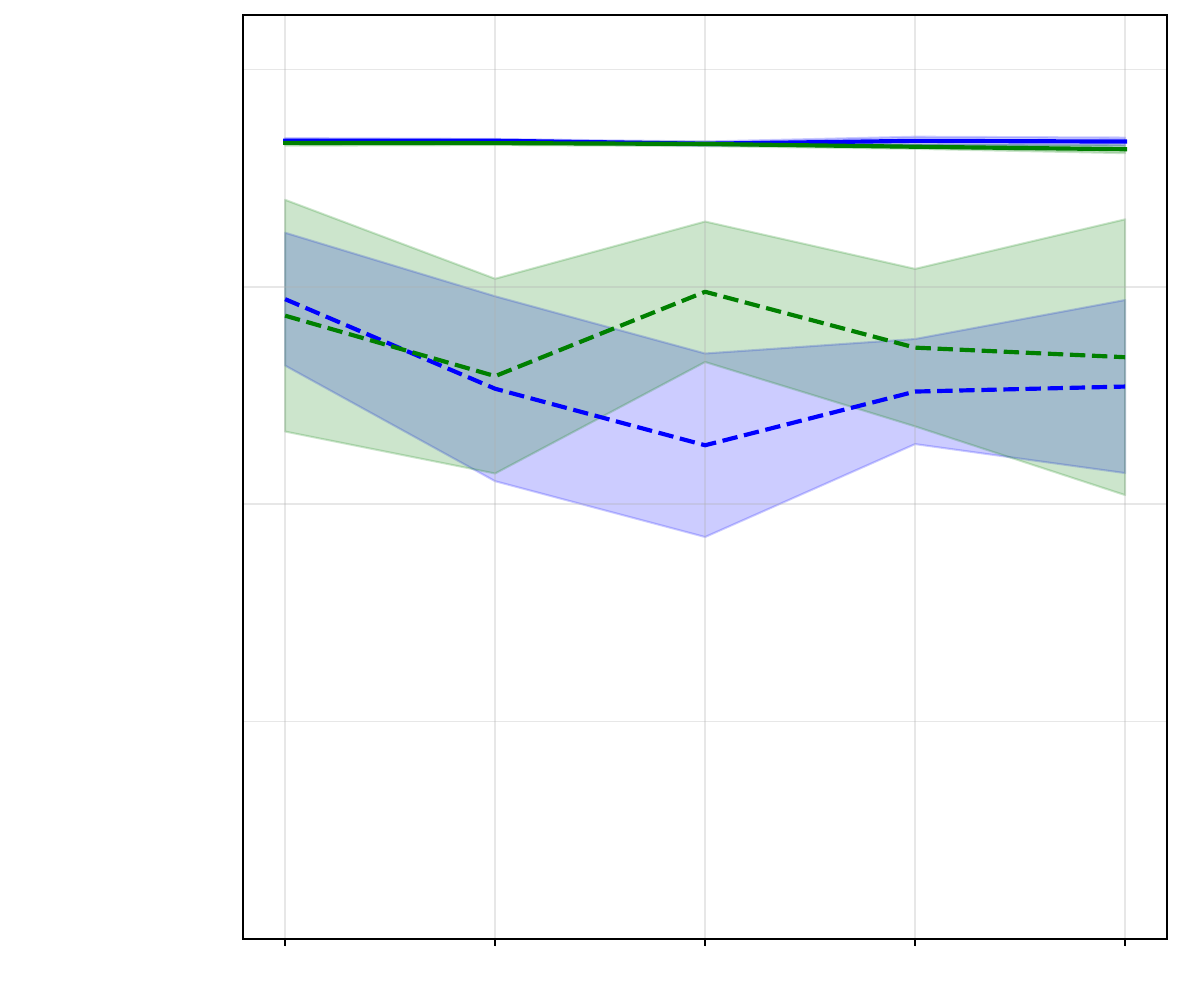}
    \caption{Deer-Horse}
\end{subfigure} 
\hspace{0.02\textwidth}
\begin{subfigure}{0.185\textwidth}
    \includegraphics[width=\linewidth]{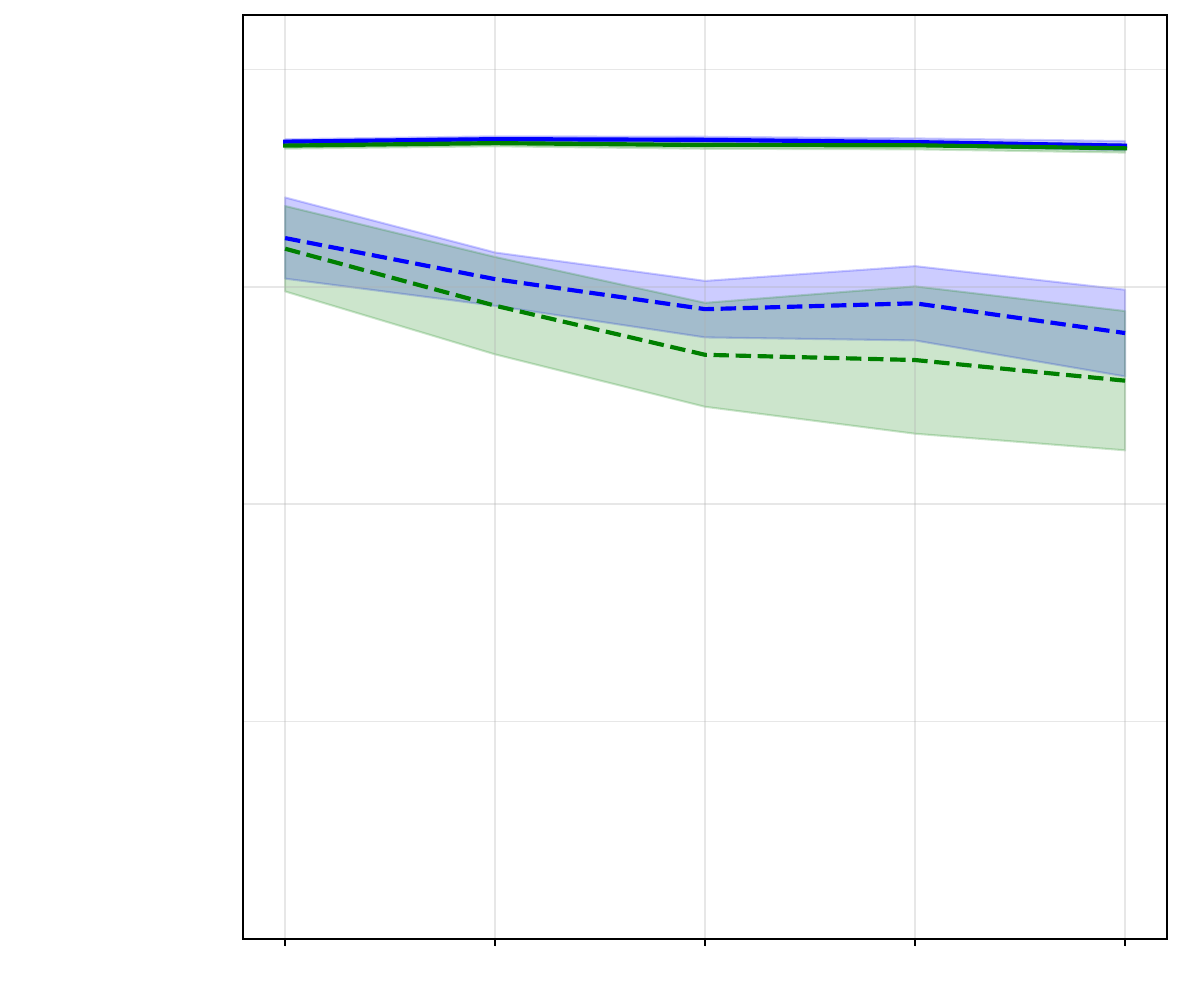}
    \caption{Ship-Truck}
\end{subfigure} 

\begin{subfigure}{0.185\textwidth}
    \includegraphics[width=\linewidth]{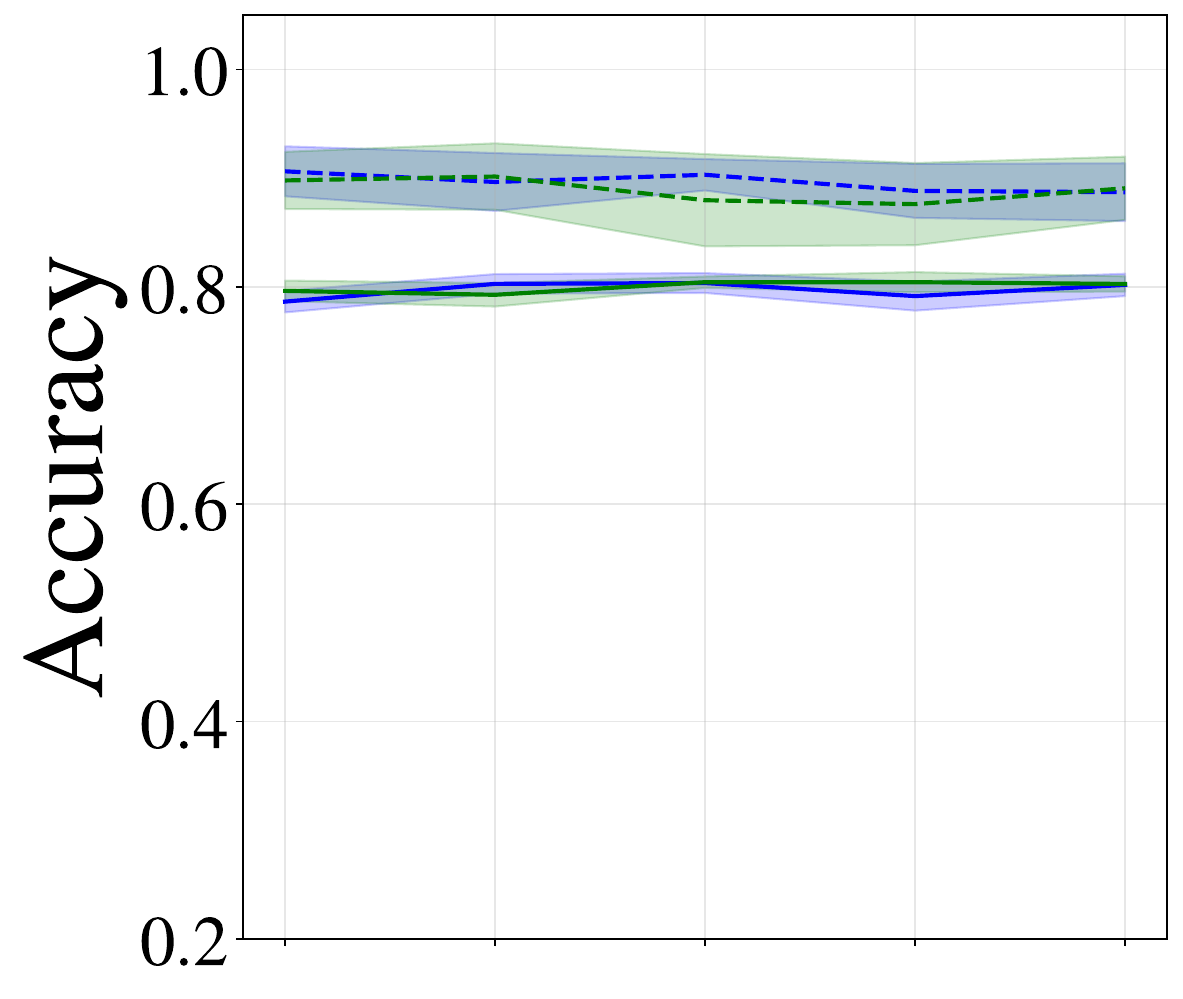}
    \caption{NV-Vasc}
\end{subfigure} 
\hspace{0.02\textwidth}
\begin{subfigure}{0.185\textwidth}
    \includegraphics[width=\linewidth]{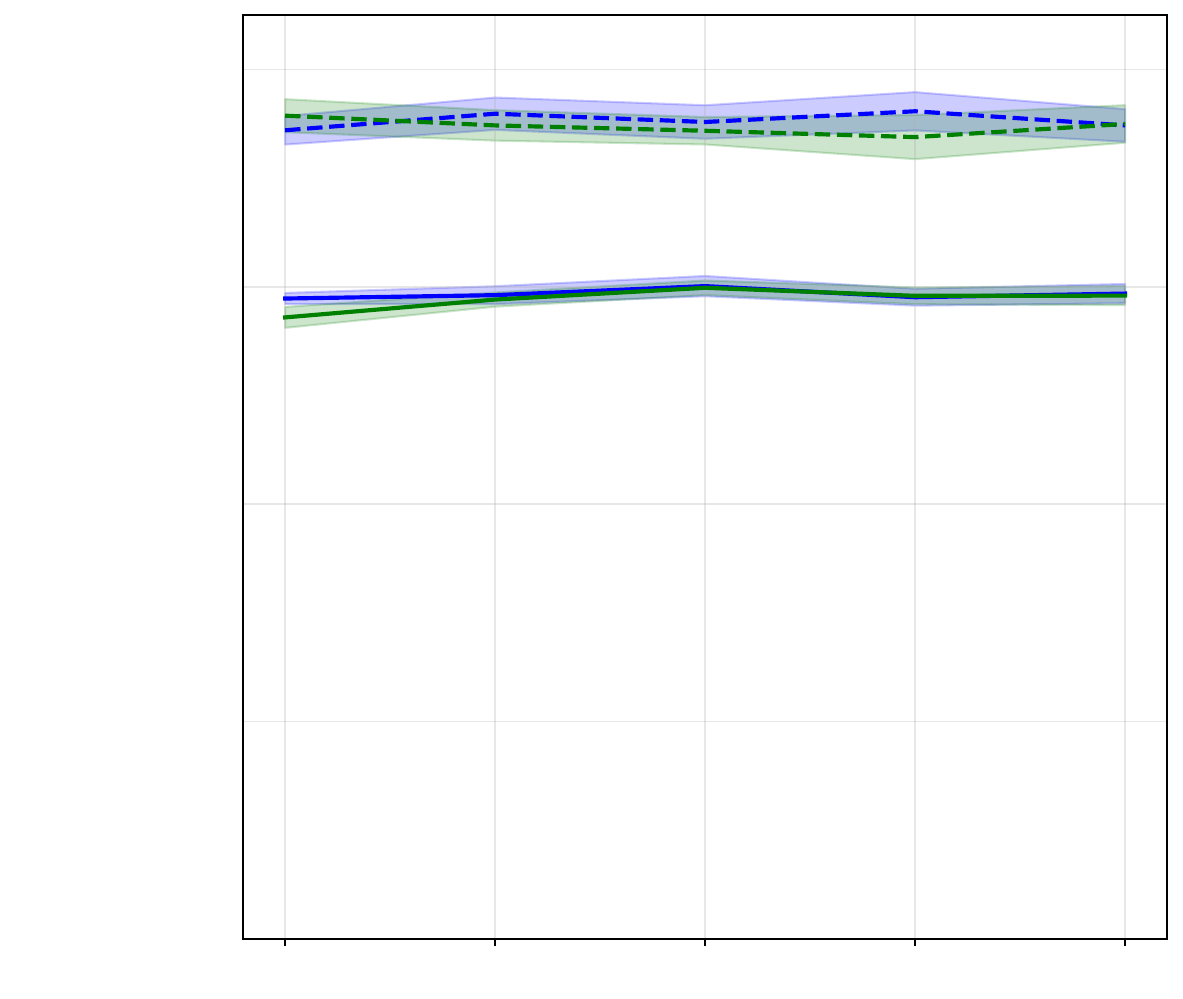}
    \caption{Mel-DF}
\end{subfigure} 
\hspace{0.02\textwidth}
\begin{subfigure}{0.185\textwidth}
    \includegraphics[width=\linewidth]{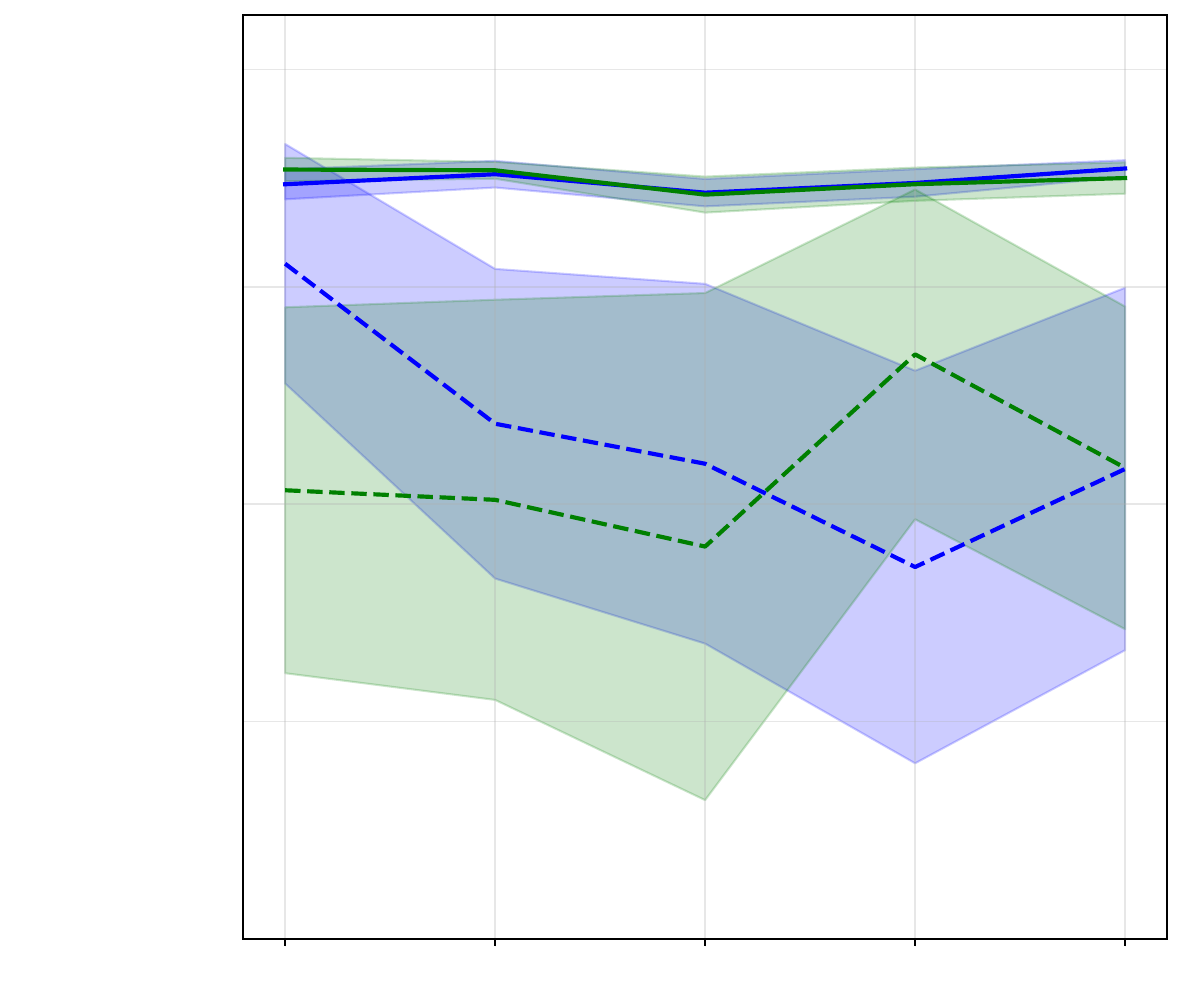}
    \caption{BG-Deb}
\end{subfigure} 

\begin{subfigure}{0.185\textwidth}
    \includegraphics[width=\linewidth]{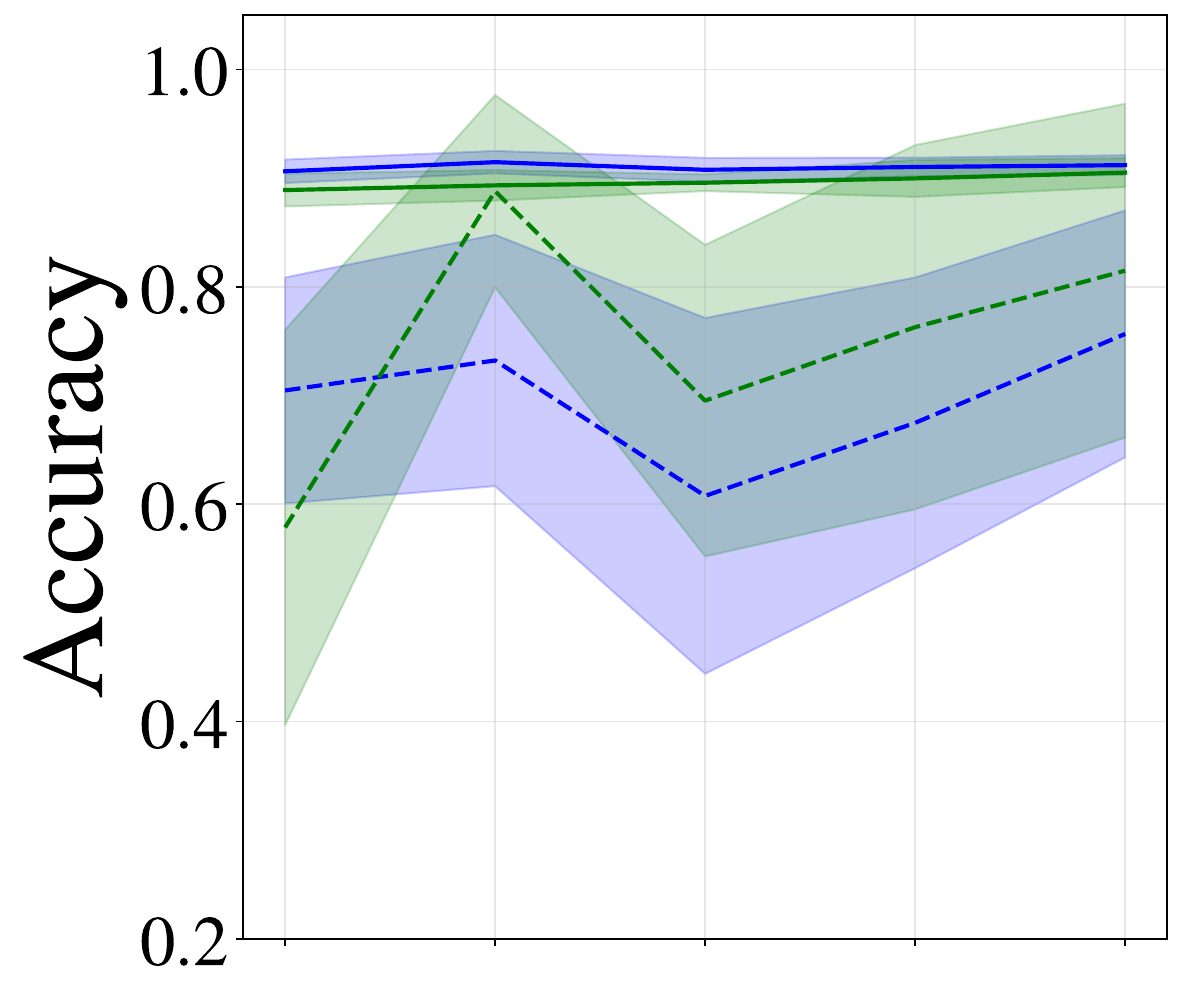}
    \caption{LymP-Deb}
\end{subfigure} 
\hspace{0.02\textwidth}
\begin{subfigure}{0.185\textwidth}
    \includegraphics[width=\linewidth]{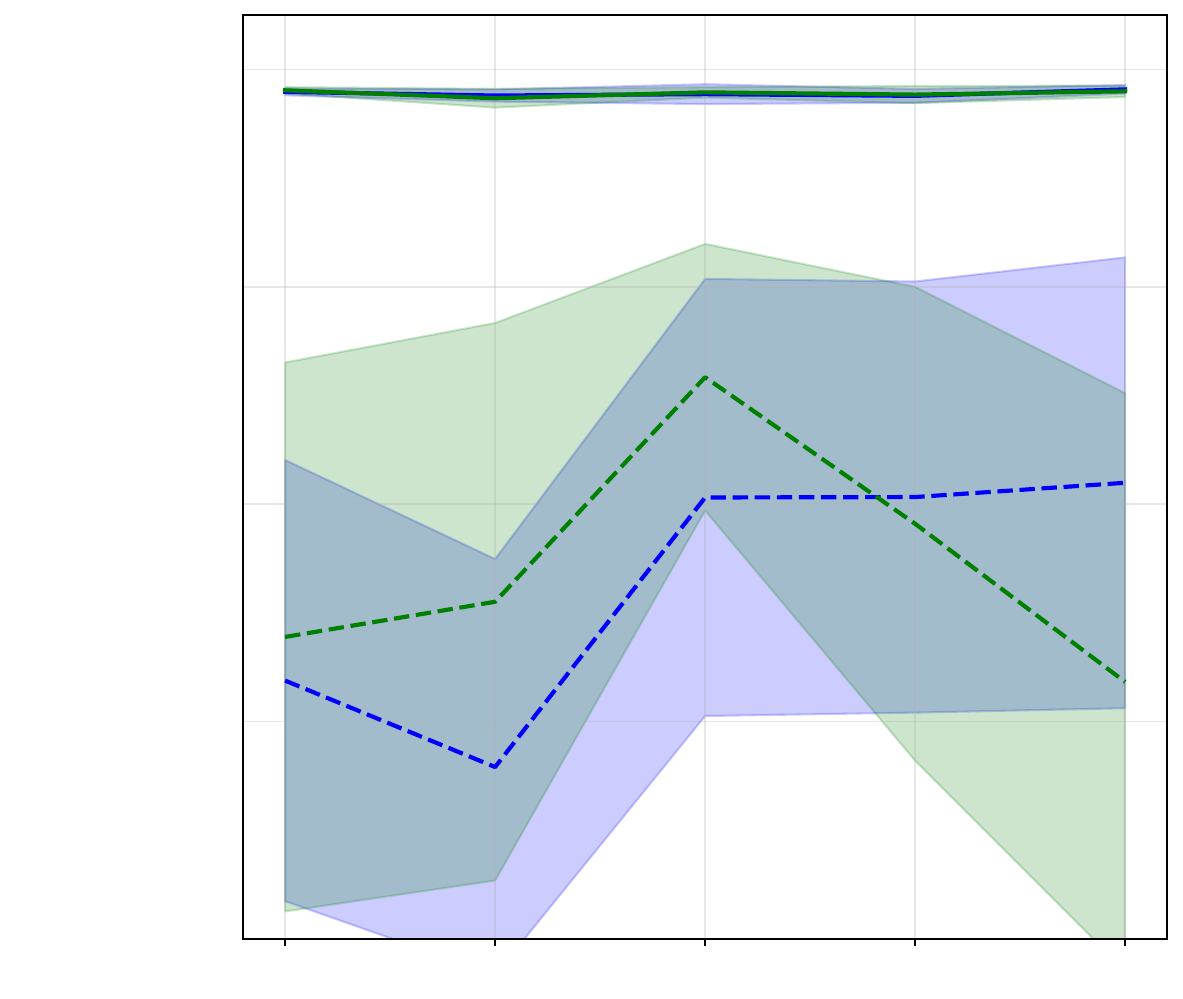}
    \caption{Baso-Eos}
\end{subfigure} 
\hspace{0.02\textwidth}
\begin{subfigure}{0.185\textwidth}
    \includegraphics[width=\linewidth]{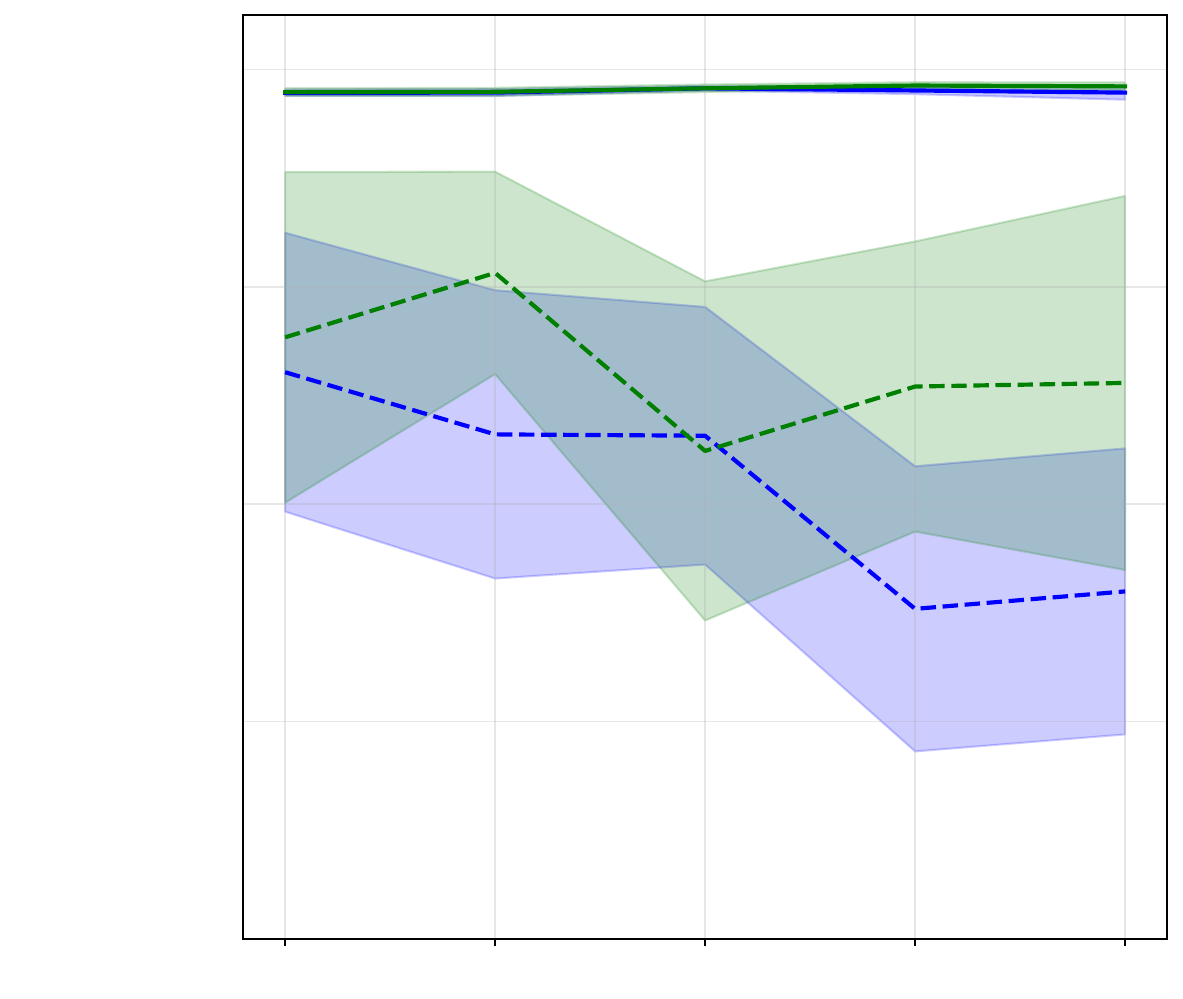}
    \caption{LymB-Mono}
\end{subfigure} 

\begin{subfigure}{0.185\textwidth}
    \includegraphics[width=\linewidth]{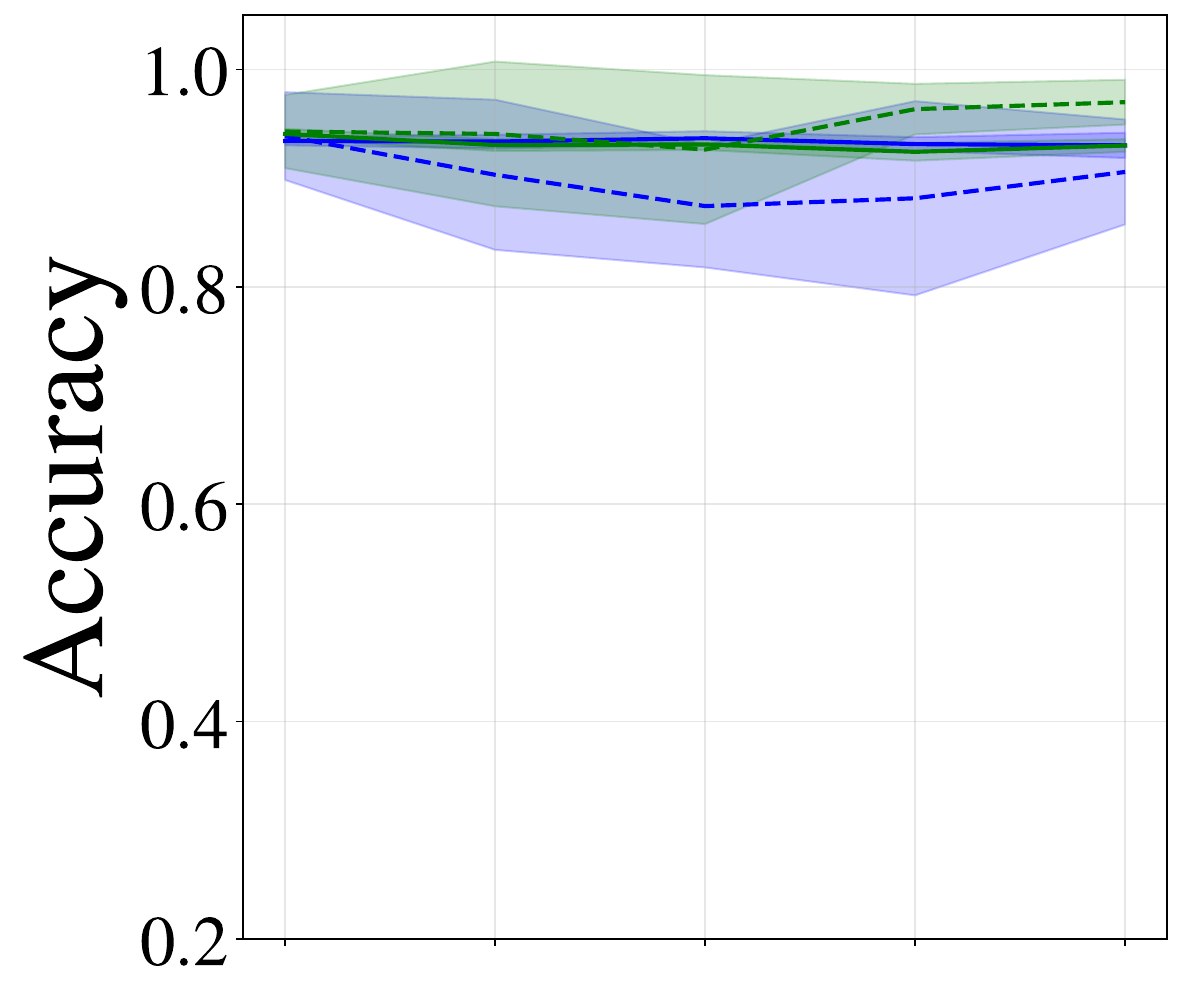}
    \caption{Auto-Truck}
\end{subfigure} 
\hspace{0.02\textwidth}
\begin{subfigure}{0.185\textwidth}
    \includegraphics[width=\linewidth]{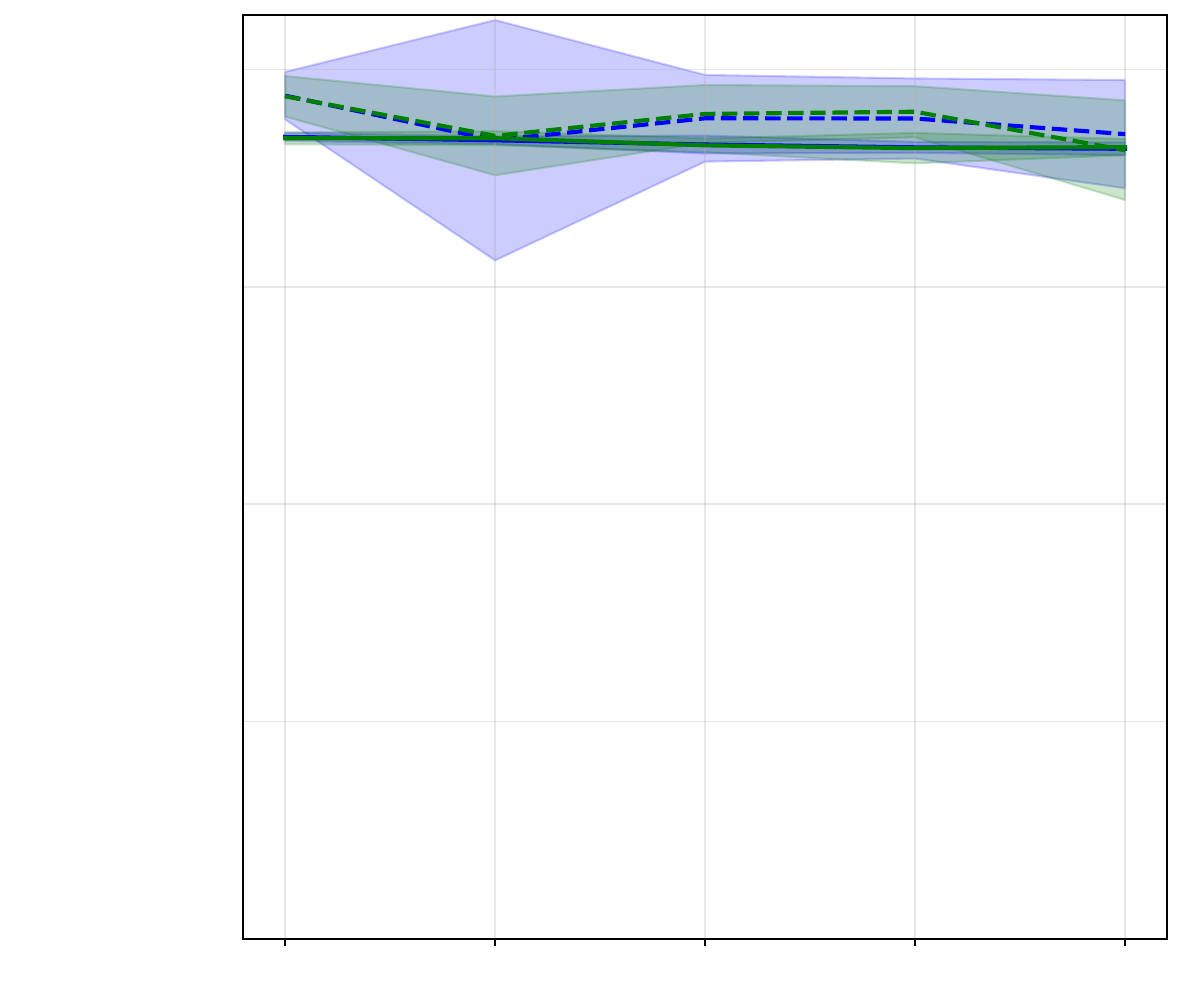}
    \caption{Truck-Auto}
\end{subfigure} 
\hspace{0.02\textwidth}
\begin{subfigure}{0.185\textwidth}
    \includegraphics[width=\linewidth]{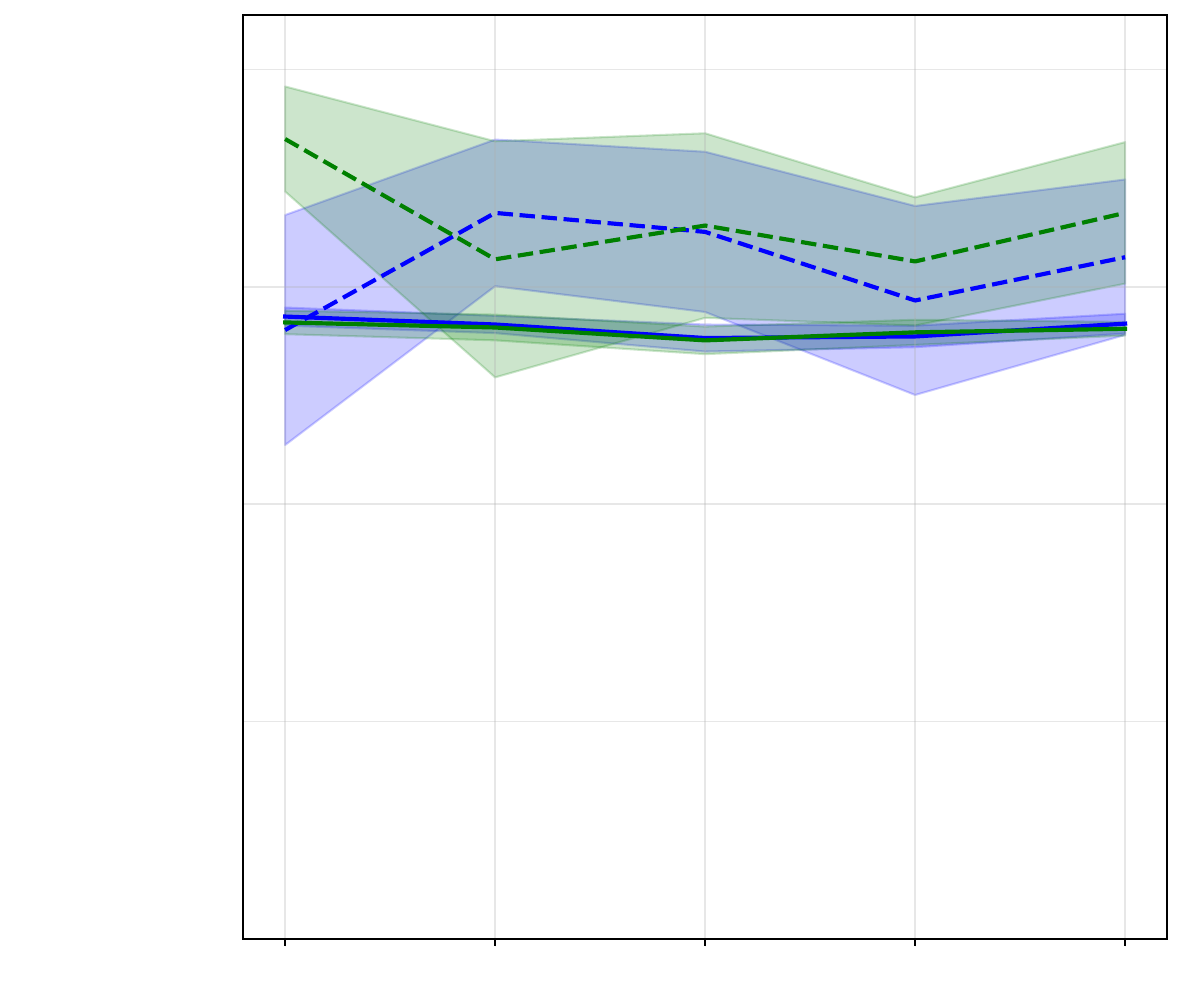}
    \caption{BCC-Mel}
\end{subfigure} 

\begin{subfigure}{0.185\textwidth}
    \includegraphics[width=\linewidth]{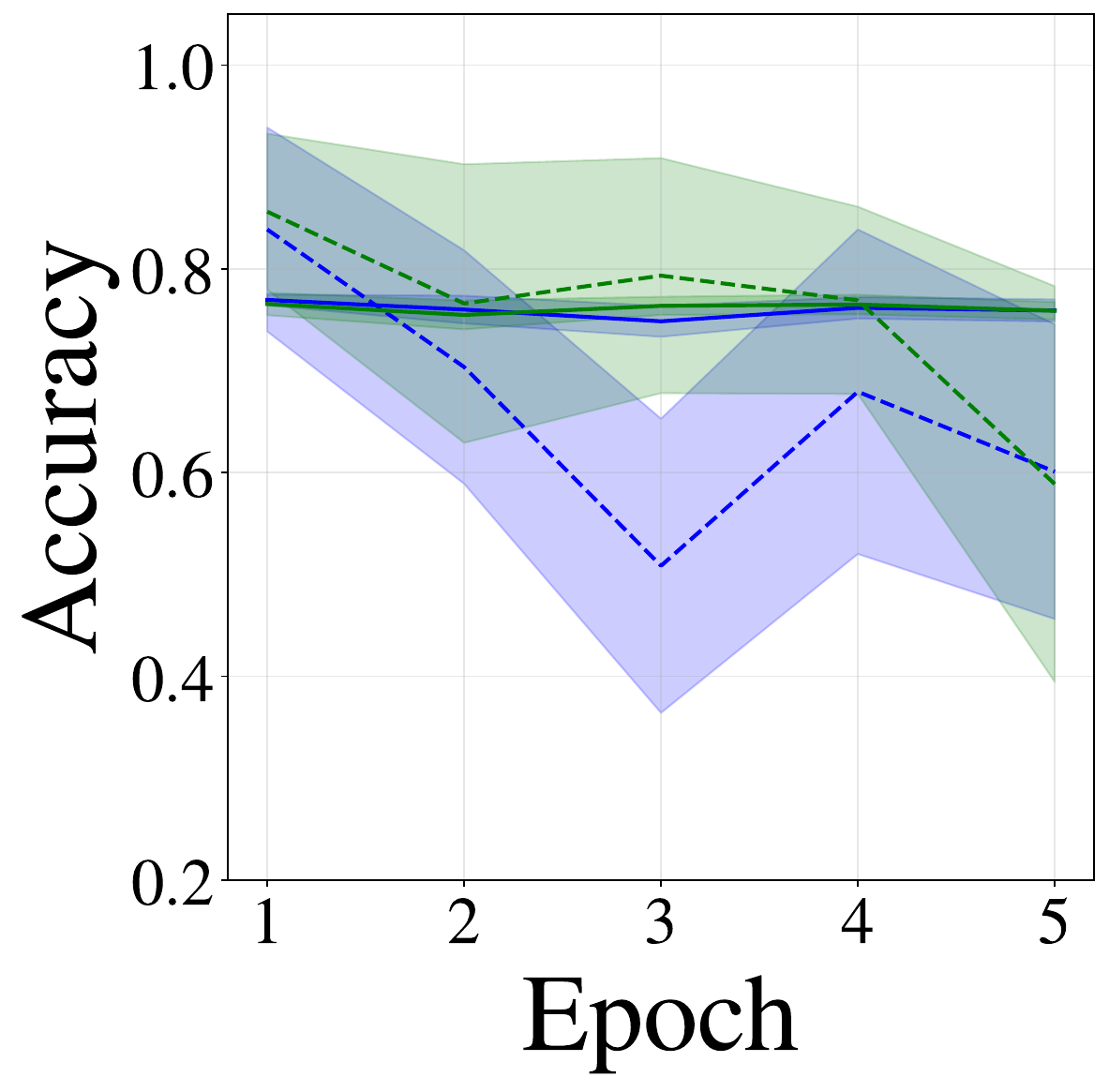}
    \caption{Mel-DF}
\end{subfigure} 
\hspace{0.02\textwidth}
\begin{subfigure}{0.185\textwidth}
    \includegraphics[width=\linewidth]{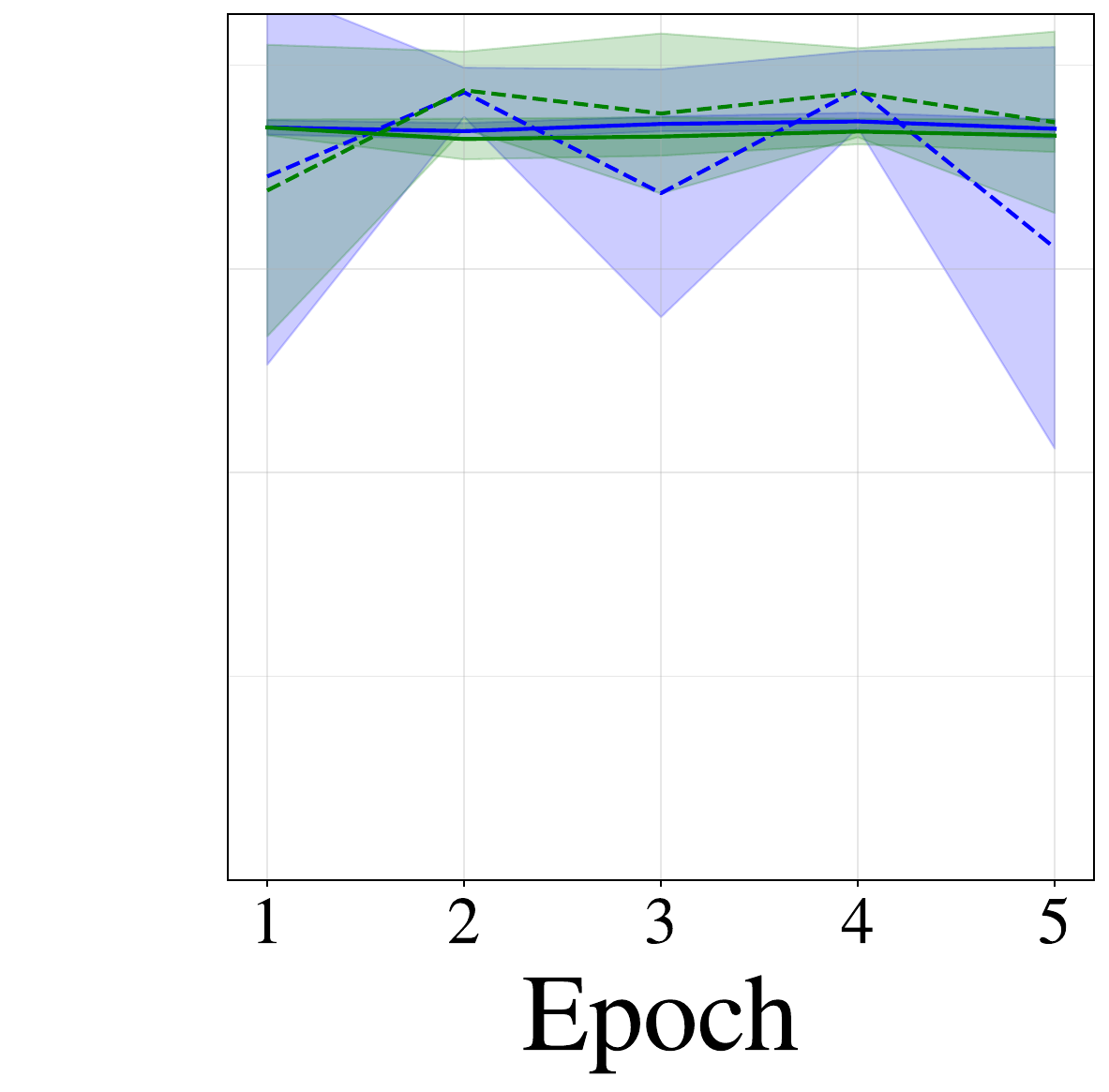}
    \caption{Adip-BG}
\end{subfigure}
\hspace{0.02\textwidth}
\begin{subfigure}{0.185\textwidth}
    \includegraphics[width=\linewidth]{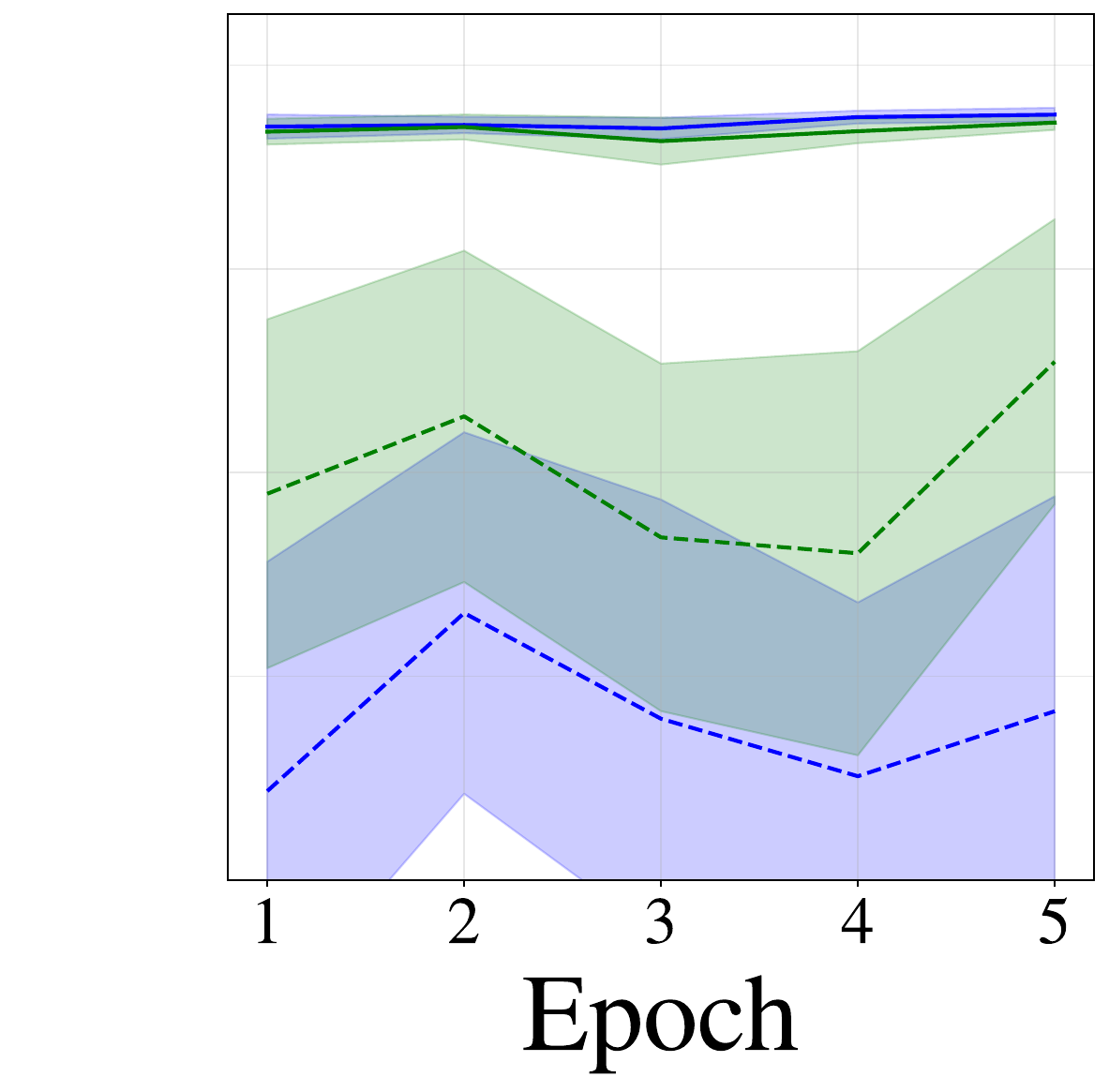}
    \caption{LymP-Deb}
\end{subfigure}

\caption{Accuracy over five epochs of (blue) fine-tuning and (green) fine-pruning defences for backdoored (a-g) ResNet18 and (h-o) ViT across four datasets. Results are reported for both clean inputs from all classes and triggered inputs from the source class. Shaded uncertainty bands represent the bounds of the 95\% confidence interval. The full class names corresponding to the abbreviations are listed in App.~\ref{app:label_abbreviations}. }
\label{fig:additional_results_tuning_defence}

\end{figure}

Figure \ref{fig:additional_results_tuning_defence} shows results for the fine-tuning and fine-pruning defences for source-target class pairs corresponding to those reported in Sec.~\ref{sec:results_analysis}. 

\FloatBarrier

\FloatBarrier

\newpage
\section{Performance of detection method SmoothInv}
\label{app:smoothinv}

\begin{figure}[htbp]
    \centering

    \begin{subfigure}[b]{0.23\textwidth}
        \centering
        \includegraphics[width=\textwidth]{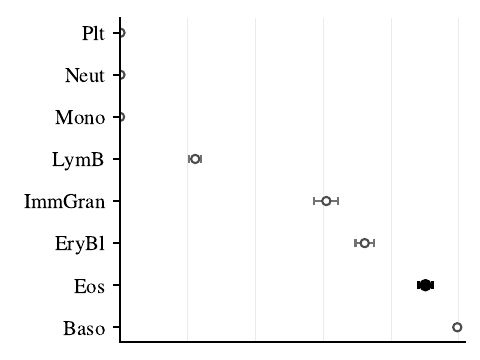}
        \caption{Baos-Eos}
        \label{fig:smoothinv_blood_resnet_basophil_eosinophil}
    \end{subfigure}
    \hfill
    \begin{subfigure}[b]{0.23\textwidth}
        \centering
        \includegraphics[width=\textwidth]{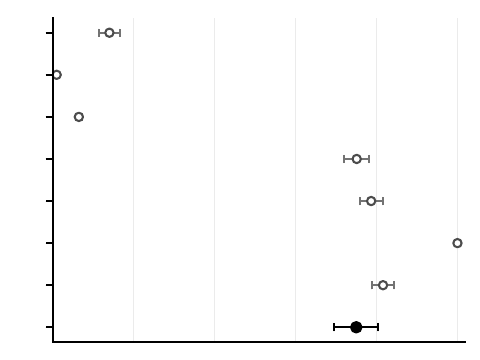}
        \caption{REryBl-Baso}
        \label{fig:smoothinv_blood_resnet_erythroblast_basophil}
    \end{subfigure}
    \hfill
    \begin{subfigure}[b]{0.23\textwidth}
        \centering
        \includegraphics[width=\textwidth]{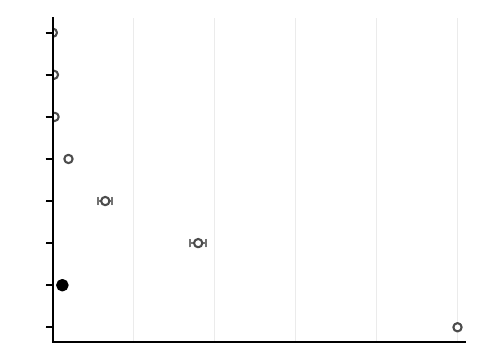}
        \caption{Baso-Eos}
        \label{fig:smoothinv_blood_vit_basophil_eosinophil}
    \end{subfigure}
    \hfill
    \begin{subfigure}[b]{0.23\textwidth}
        \centering
        \includegraphics[width=\textwidth]{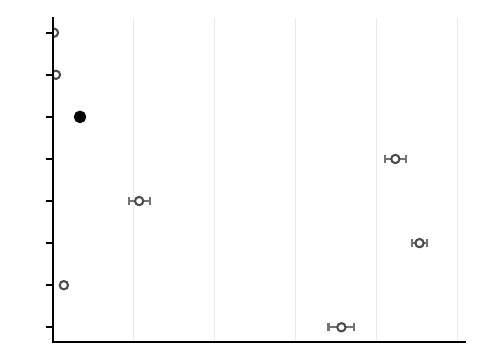}
        \caption{LymB-Mono}
        \label{fig:smoothinv_blood_vit_lymphocyte_monocyte}
    \end{subfigure}

    \vspace{0.5em}

    \begin{subfigure}[b]{0.23\textwidth}
        \centering
        \includegraphics[width=\textwidth]{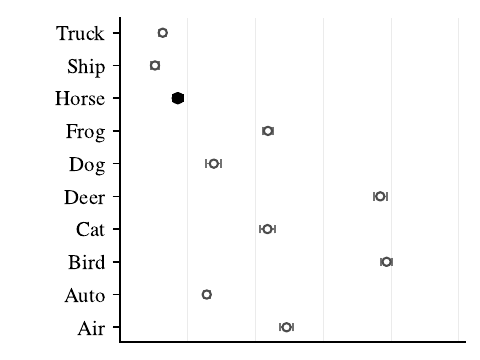}
        \caption{Deer-Horse}
        \label{fig:smoothinv_cifar_resnet_deer_horse}
    \end{subfigure}
    \hfill
    \begin{subfigure}[b]{0.23\textwidth}
        \centering
        \includegraphics[width=\textwidth]{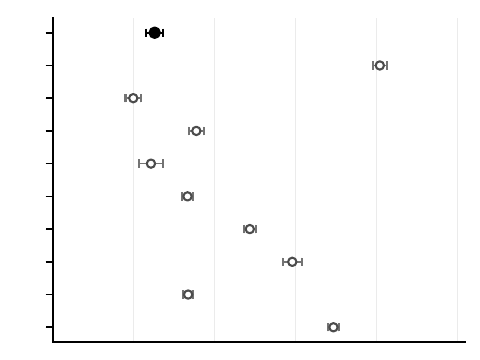}
        \caption{Ship-Truck}
        \label{fig:smoothinv_cifar_resnet_ship_truck}
    \end{subfigure}
    \hfill
    \begin{subfigure}[b]{0.23\textwidth}
        \centering
        \includegraphics[width=\textwidth]{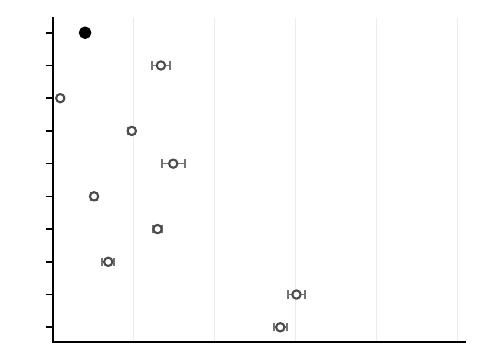}
        \caption{Auto-Truck}
        \label{fig:smoothinv_cifar_vit_automobile_truck}
    \end{subfigure}
    \hfill
    \begin{subfigure}[b]{0.23\textwidth}
        \centering
        \includegraphics[width=\textwidth]{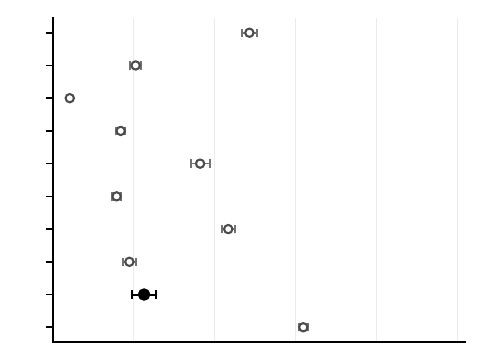}
        \caption{Truck-Auto}
        \label{fig:smoothinv_cifar_vit_truck_automobile}
    \end{subfigure}

    \vspace{0.5em}

    \begin{subfigure}[b]{0.23\textwidth}
        \centering
        \includegraphics[width=\textwidth]{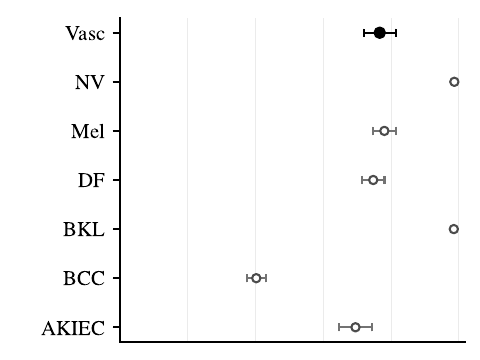}
        \caption{NV-Vasc}
        \label{fig:smoothinv_derma_resnet_melanocytic_vascular}
    \end{subfigure}
    \hfill
    \begin{subfigure}[b]{0.23\textwidth}
        \centering
        \includegraphics[width=\textwidth]{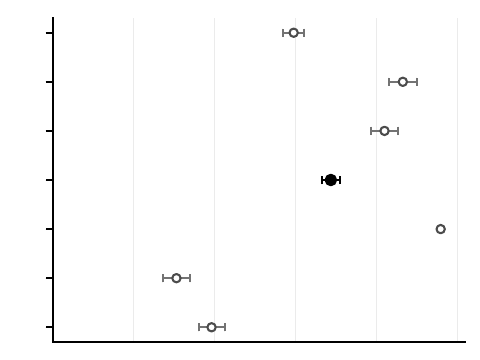}
        \caption{Mel-DF}
        \label{fig:smoothinv_derma_resnet_melanoma_dermatofibroma}
    \end{subfigure}
    \hfill
    \begin{subfigure}[b]{0.23\textwidth}
        \centering
        \includegraphics[width=\textwidth]{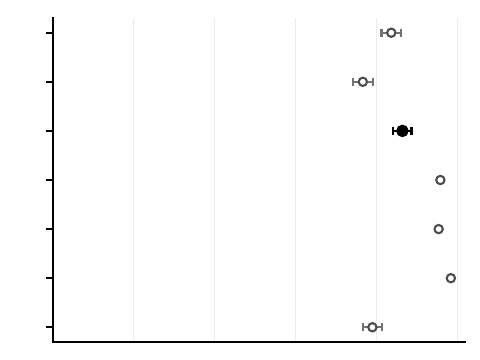}
        \caption{BCC-Mel}
        \label{fig:smoothinv_derma_vit_basal_melanoma}
    \end{subfigure}
    \hfill
    \begin{subfigure}[b]{0.23\textwidth}
        \centering
        \includegraphics[width=\textwidth]{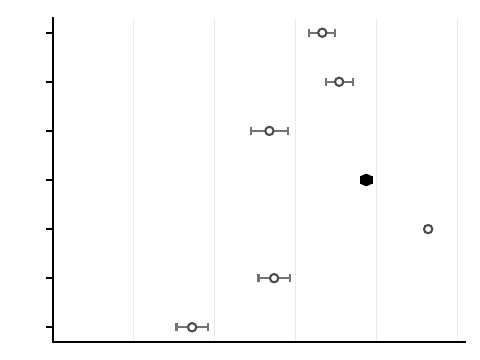}
        \caption{Mel-DF}
        \label{fig:smoothinv_derma_vit_melanoma_dermatofibroma}
    \end{subfigure}

    \vspace{0.5em}

    \begin{subfigure}[b]{0.23\textwidth}
        \centering
        \includegraphics[width=\textwidth]{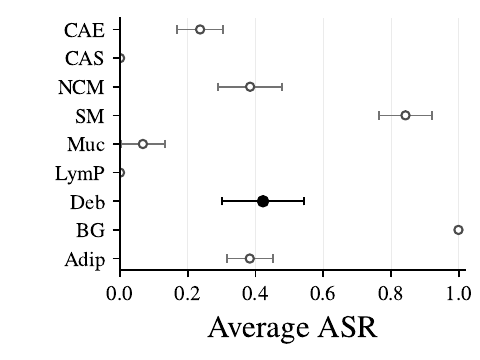}
        \caption{BG-Deb}
        \label{fig:smoothinv_path_resnet_background_debris}
    \end{subfigure}
    \hfill
    \begin{subfigure}[b]{0.23\textwidth}
        \centering
        \includegraphics[width=\textwidth]{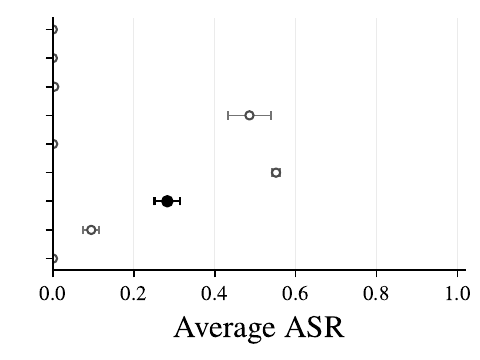}
        \caption{LymP-Deb}
        \label{fig:smoothinv_path_resnet_lymphocytes_debris}
    \end{subfigure}
    \hfill
    \begin{subfigure}[b]{0.23\textwidth}
        \centering
        \includegraphics[width=\textwidth]{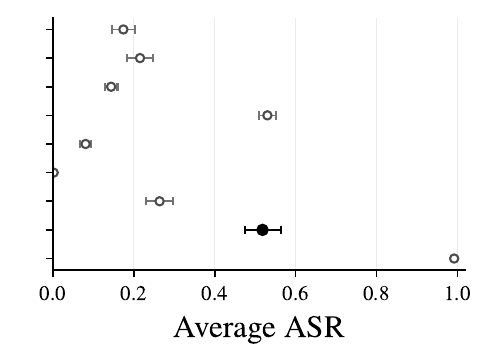}
        \caption{Adip-BG}
        \label{fig:smoothinv_path_vit_adipose_background}
    \end{subfigure}
    \hfill
    \begin{subfigure}[b]{0.23\textwidth}
        \centering
        \includegraphics[width=\textwidth]{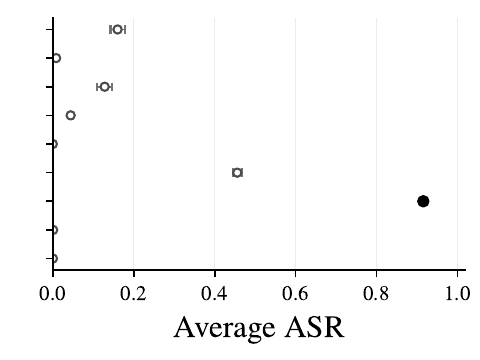}
        \caption{LymP-Deb}
        \label{fig:smoothinv_path_vit_lymphocytes_debris}
    \end{subfigure}
    \caption{SmoothInv results for backdoored (a, b, c, f, i, j, m, n) ResNet18 and (c, d, g, h, k, l, o, p) ViT across four datasets. Results are reported as the average ASR per class with associated 95\% confidence intervals. The full class names corresponding to the abbreviations are listed in App.~\ref{app:label_abbreviations}. }
    \label{fig:smoothinv_score_profile}
\end{figure}

The results for the detection method SmoothInv \citep{sun2023single} are reported as the mean ASR of the optimised SmoothInv trigger over 10 random seeds, with 95\% confidence intervals. Figure~\ref{fig:smoothinv_score_profile} shows the ASR profile across classes, with the black marker indicating the true target class. Successful identification should result in a clear separation between the true target and the remaining classes. 

We use perturbation size $\epsilon = 10$ and no diffusion, see details in \citet{sun2023single}. Notably, we only test using source-class images; however, the defender does not know that the attack is single-class or which source class is affected, and would therefore need to evaluate all source-target combinations (or the entire test set), substantially increasing the computational cost. We consider SmoothInv successful if the recovered trigger yields a high ASR for the true target class while producing substantially lower ASR for non-target classes.

The results indicate that SmoothInv does not reliably identify the true backdoored target class across configurations, producing high ASR also for non-target classes, which suggests limited specificity (high false positive rate).

\FloatBarrier
\section{Hyperparameter sweeps and sensitivity analysis}
\label{app:hyperparameter-sweeps}

To illustrate the trade-offs between the key hyperparameters -- the spike strength $\theta$ and stealthiness, measured by the regularisation weight $\lambda_\text{L2}$ -- and their effect on ASR, we perform separate parameter sweeps while keeping all other hyperparameters fixed at the values reported in App.~\ref{app:hyperparameter_config}.

Table~\ref{tab:asr_clean_theta} reports the ASR and clean accuracy of backdoors planted with different values of the spike strength $\theta$. As expected, making the spike more prominent by increasing $\theta$ increases the ASR but decreases clean accuracy: a stronger spike has a greater effect on triggered inputs, but its dominance in the model weights reduces performance on clean inputs. Table~\ref{tab:asr_vs_l2} shows the effect of varying the stealthiness through the regularisation weight $\lambda_\text{L2}$. As expected, smaller values of $\lambda_\text{L2}$ impose weaker regularisation on the optimised trigger and result in higher ASR. Conversely, stronger regularisation produces less visible perturbations and a triggered input that more closely resembles clean images, leading to lower ASR. 

Both experiments use the ResNet18 architecture trained on the BloodMNIST dataset with source class \textit{basophil} and target class \textit{eosinophil}. Each experiment is run with 10 random seeds, and results are reported with 95\% confidence intervals.

\begin{table}[H]
\centering

\begin{minipage}[t]{0.46\textwidth}
\centering
\caption{Average ASR and clean accuracy for different spike strengths $\theta$, with 95\% confidence intervals.}
\label{tab:asr_clean_theta}
\begin{tabular}{rll}
\toprule
$\theta$ & ASR & Clean Accuracy \\
\midrule
0.010 & 0.239 $\pm$ 0.043 & 0.982 $\pm$ 0.000 \\
0.025 & 0.881 $\pm$ 0.029 & 0.982 $\pm$ 0.000 \\
0.050 & 0.940 $\pm$ 0.011 & 0.982 $\pm$ 0.000 \\
0.100 & 0.969 $\pm$ 0.007 & 0.982 $\pm$ 0.000 \\
0.250 & 0.986 $\pm$ 0.004 & 0.974 $\pm$ 0.002 \\
0.500 & 0.999 $\pm$ 0.002 & 0.923 $\pm$ 0.006 \\
1.000 & 1.000 $\pm$ 0.000 & 0.807 $\pm$ 0.008 \\
\bottomrule
\end{tabular}
\end{minipage}
\hfill
\begin{minipage}[t]{0.46\textwidth}
\centering
\caption{Average ASR for different regularisation weights $\lambda_{\text{L2}}$, with 95\% confidence intervals.}
\label{tab:asr_vs_l2}
\begin{tabular}{rl}
\toprule
$\lambda_{\text{L2}}$ & ASR \\
\midrule
0 & 1.000 $\pm$ 0.000 \\
5 & 1.000 $\pm$ 0.000 \\
10 & 0.991 $\pm$ 0.003 \\
15 & 0.964 $\pm$ 0.009 \\
20 & 0.936 $\pm$ 0.007 \\
25 & 0.789 $\pm$ 0.081 \\
30 & 0.352 $\pm$ 0.043 \\
\bottomrule
\end{tabular}
\end{minipage}

\end{table}

\FloatBarrier

\section{Visual stealthiness of input trigger and the Maximum Mean Discrepancy metric}
\label{app:mmd}

\begin{figure}[h]
    \centering

    \begin{subfigure}{0.45\textwidth}
        \centering
        \includegraphics[width=\linewidth]{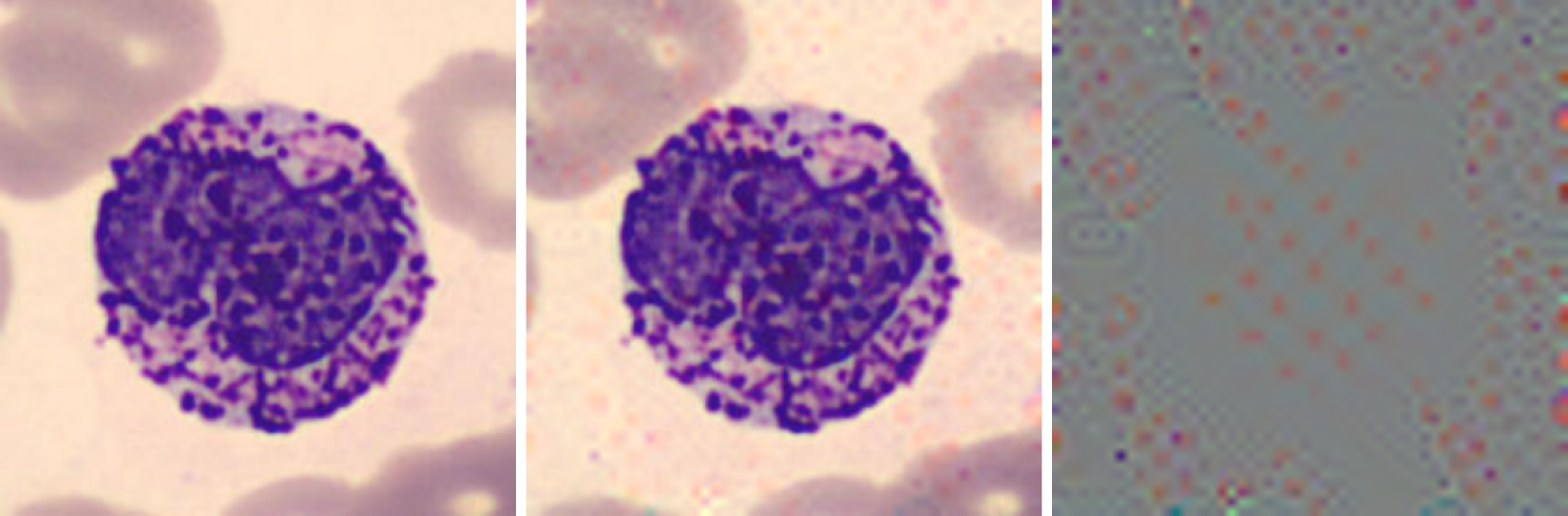}
        \caption{}
    \end{subfigure}
    \hfill
    \begin{subfigure}{0.45\textwidth}
        \centering
        \includegraphics[width=\linewidth]{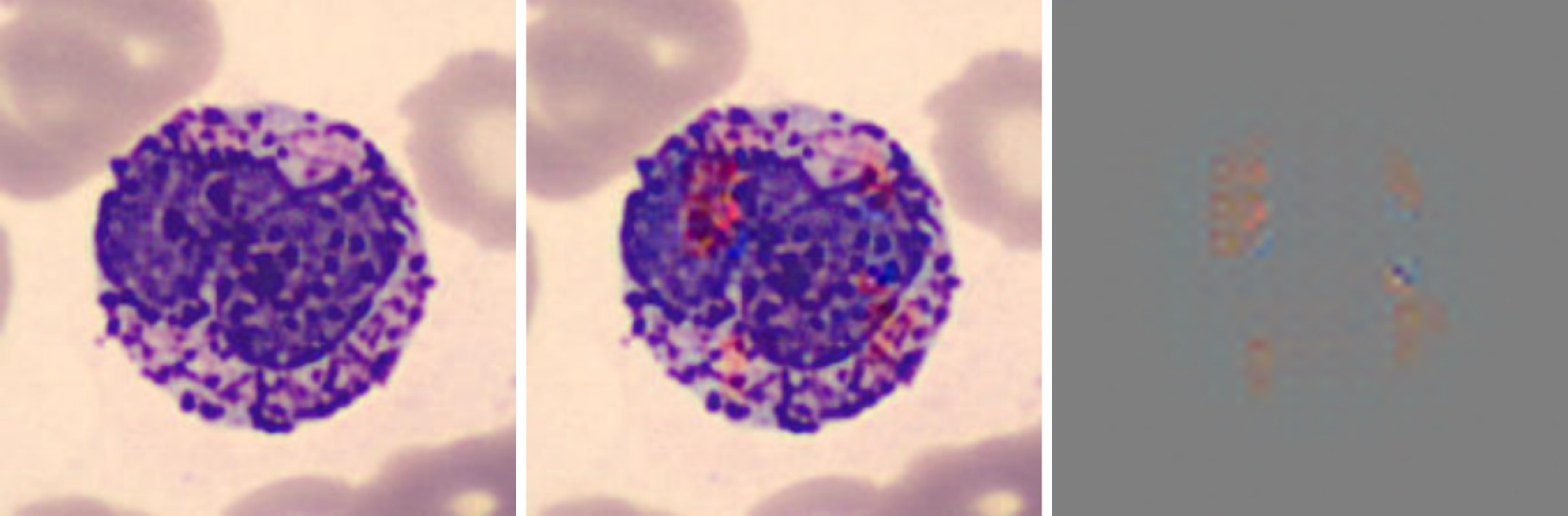}
        \caption{}
    \end{subfigure}

    \vspace{0.5cm}

    \begin{subfigure}{0.45\textwidth}
        \centering
        \includegraphics[width=\linewidth]{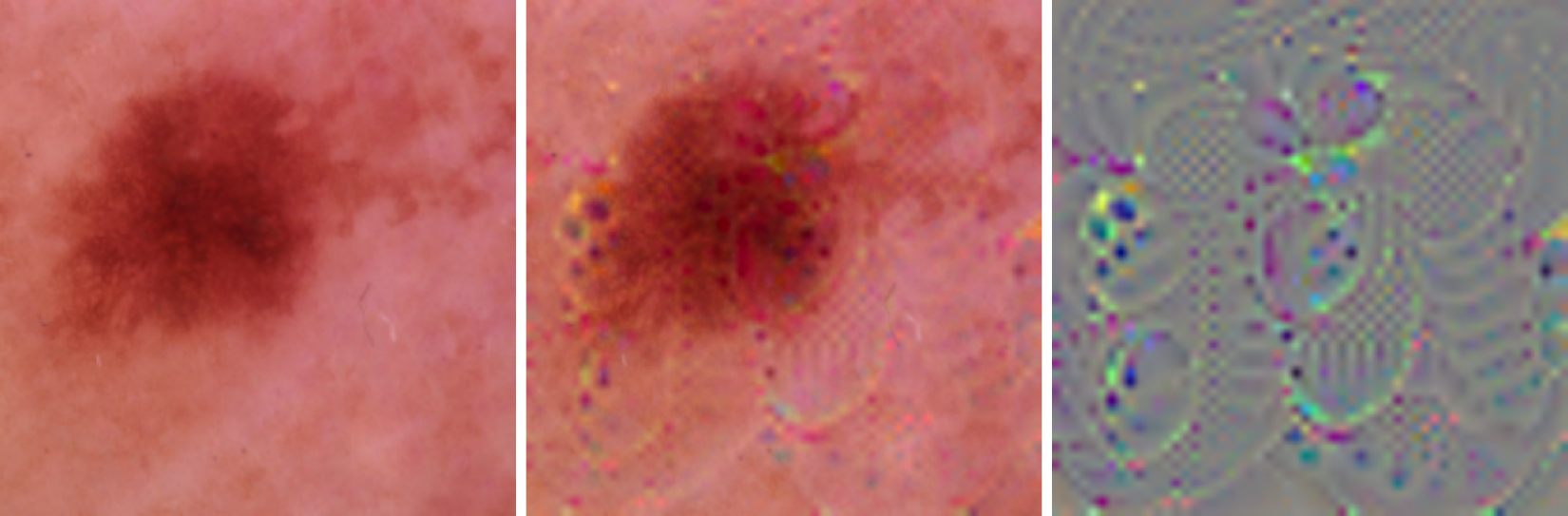}
        \caption{}
    \end{subfigure}
    \hfill
    \begin{subfigure}{0.45\textwidth}
        \centering
        \includegraphics[width=\linewidth]{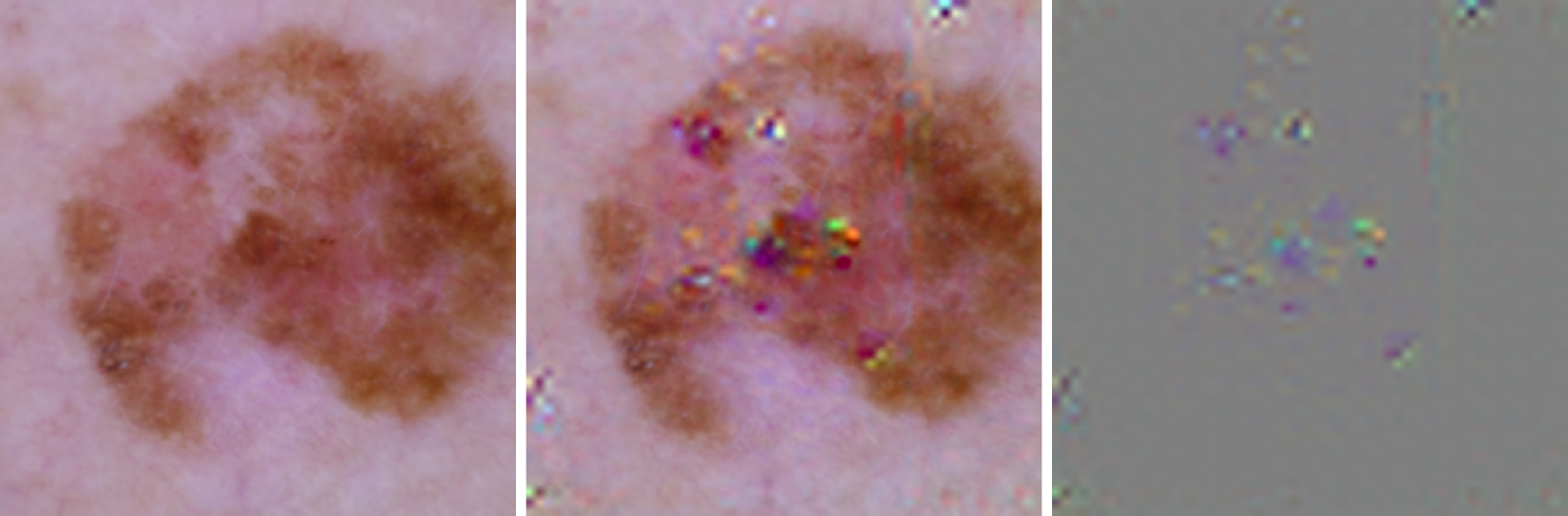}
        \caption{}
    \end{subfigure}

    \caption{Original source class (left), triggered input (middle), and secret key (right, shown on a neutral background) images for different model-dataset configurations: (a) ResNet18 on BloodMNIST with source class \textit{basophil} and target class \textit{eosinophil}, (b) ViT on the same dataset and class pair, (c) ResNet18 on DermaMNIST with source class \textit{melanocytic nevi} and target class \textit{vascular lesions}, and (d) ViT on DermaMNIST with source class \textit{melanoma} and target class \textit{dermatofibroma}. Images are randomly selected and illustrate the visual effect of applying the secret key with scaling factor $\lambda$ as used in experiments (Sec.~\ref{sec:results_analysis}). 
    }
    \label{fig:secret_key_in}
\end{figure}

\begin{figure*}[h]
    \centering

    \begin{subfigure}[b]{0.55\textwidth}
        \centering
        \includegraphics[width=\linewidth]{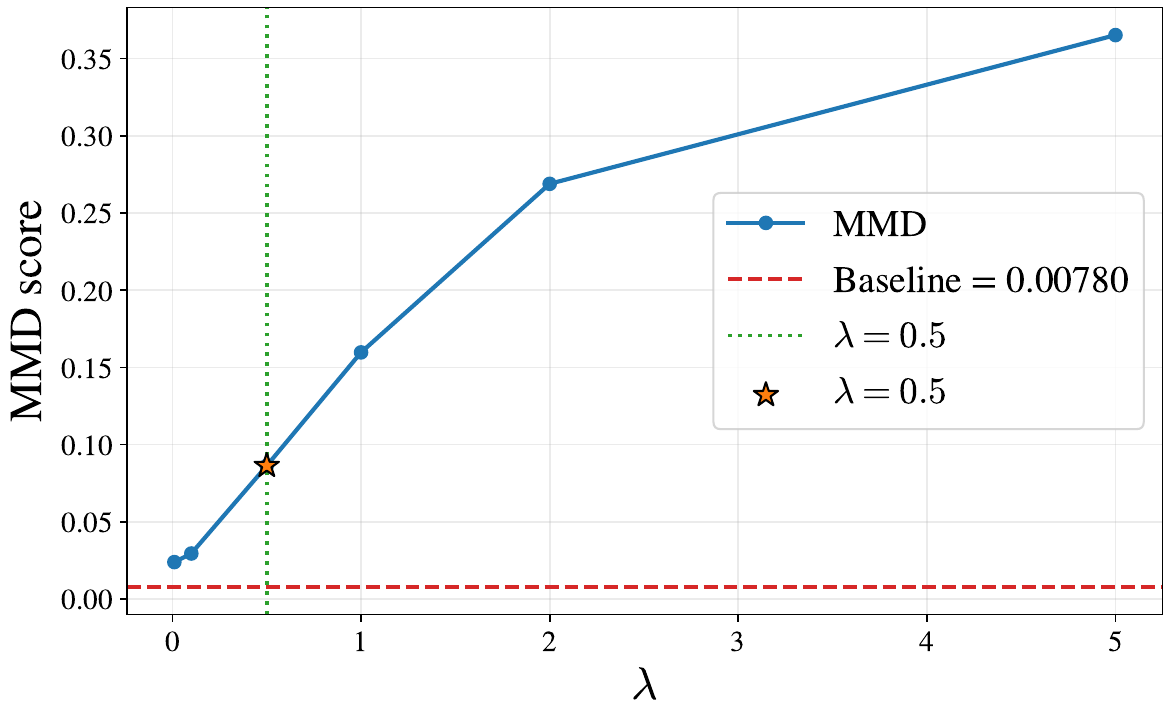}
        \caption{}
    \end{subfigure}
    \hfill
    \begin{subfigure}[b]{0.4\textwidth}
        \centering
        
        \begin{subfigure}{0.35\linewidth}
            \centering
            \includegraphics[width=\linewidth]{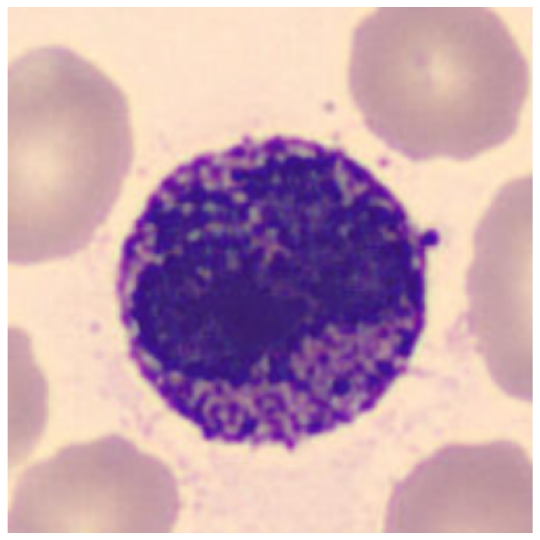}
     
        \end{subfigure}
        \hspace{0.02\textwidth}
        \begin{subfigure}{0.35\linewidth}
            \centering
            \includegraphics[width=\linewidth]{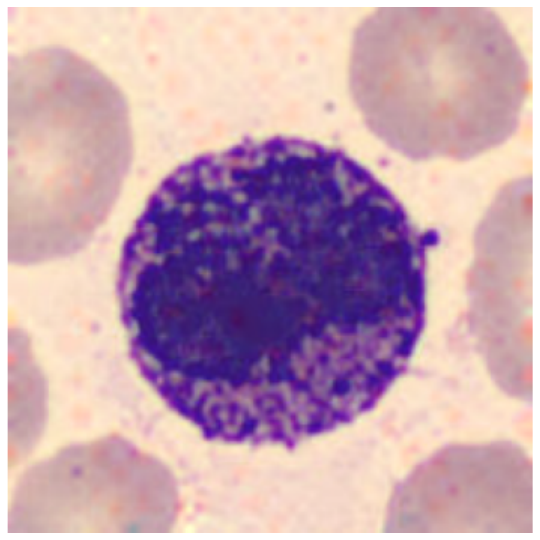}

        \end{subfigure}

        \hspace{0.02\textwidth}

        \begin{subfigure}{0.35\linewidth}
            \centering
            \includegraphics[width=\linewidth]{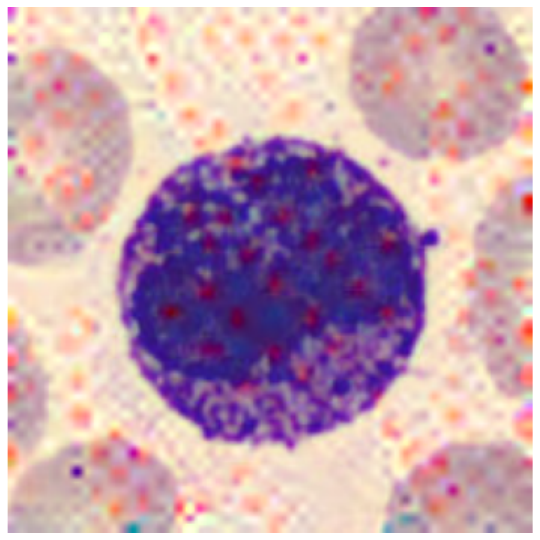}
       
        \end{subfigure}
        \hspace{0.02\textwidth}
        \begin{subfigure}{0.35\linewidth}
            \centering
            \includegraphics[width=\linewidth]{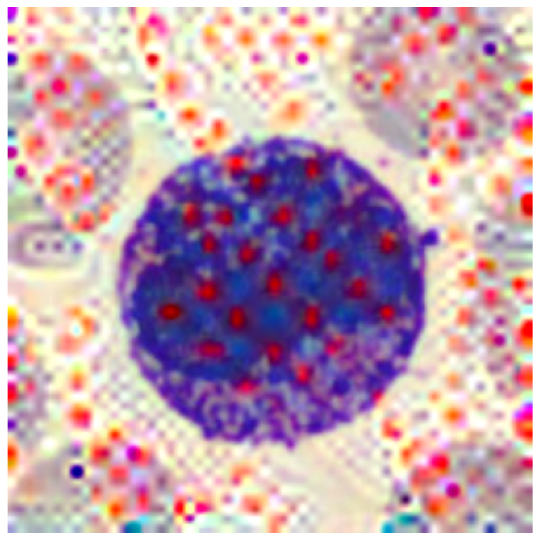}
        
        \end{subfigure}
        \vspace{12pt}
        \caption{}
    \end{subfigure}
    \caption{(a) Visibility of the trigger, quantified by the MMD score, as a function of the scaling factor $\lambda$. Results are shown for ResNet18 on BloodMNIST with source class \textit{basophil} and target class \textit{eosinophil}. The value of $\lambda$ used in the final configuration giving results as presented in Sec.~\ref{sec:results_analysis} is highlighted. A baseline MMD score is computed by splitting the original dataset into two halves. (b) Example input image from the source class with the same secret key weighted by (top left) $\lambda = 0.01$, (top right) $\lambda = 0.5$, (bottom left) $\lambda = 2$ and (bottom right) $\lambda = 5$.}
    \label{fig:mmd_vs_lambda}
\end{figure*}

Figure \ref{fig:secret_key_in} shows four examples of clean inputs alongside their triggered counterparts activating the backdoor. Figure \ref{fig:mmd_vs_lambda} illustrates the influence of the scaling factor $\lambda$ on the perceptibility of the trigger, showing that larger values of $\lambda$ yield more visible perturbations. Stealthiness in terms of visual imperceptibility is quantified by the Maximum Mean Discrepancy (MMD) score \citep{gretton2012kernel} as the discrepancy between the original images and their triggered counterparts.  

MMD is a statistical measure determining whether two sets of samples are drawn from the same underlying distribution by computing the distance between their mean embeddings. Both the original and corresponding triggered images are passed through a pretrained InceptionV3 network to extract high-level feature representations denoted as $\mathcal{X} = \{\mathbf{x}_i\}_{i=1}^{n} \subset \mathbb{R}^p$ (original) and $\mathcal{Y} = \{\mathbf{y}_j\}_{j=1}^{m} \subset \mathbb{R}^p$ (triggered). A polynomial kernel function $k : \mathbb{R}^p \times \mathbb{R}^p \rightarrow \mathbb{R}$ defined as $k(\mathbf{x}, \mathbf{y}) = (\gamma \mathbf{x}^\top \mathbf{y} + c)^d$, with scaling factor $\gamma = \frac{1}{p}$, constant offset $c = 1$ and polynomial degree $d = 3$, computes the pairwise similarities of the feature embeddings. The biased empirical estimate of MMD is given by

\begin{equation}
\begin{aligned}
\text{MMD}(\mathcal{X}, \mathcal{Y})
&= \frac{1}{n^2} \sum_{i,j} k(\mathbf{x}_i,\mathbf{x}_j)
 + \frac{1}{m^2} \sum_{i,j} k(\mathbf{y}_i,\mathbf{y}_j) \\
&\quad - \frac{2}{nm} \sum_{i,j} k(\mathbf{x}_i,\mathbf{y}_j),
\end{aligned}
\label{eq:mmd}
\end{equation}

where the terms measure intra-set similarity of original features, intra-set similarity of triggered features, and cross-set similarity between original and triggered features, respectively. If the two sets originate from the same distribution, the MMD value approaches zero. Thus, a lower MMD score indicates higher similarity between the original and triggered image distributions, while larger values suggest stronger distributional shifts introduced by the trigger. 

\FloatBarrier

\end{document}